\documentclass[11pt,twoside]{article}
\usepackage{graphicx}

\usepackage{asp2006}
\usepackage{epsf}
\usepackage{psfig}
\usepackage{lscape}

\newcommand{\bdv}[1]{\mbox{\boldmath$#1$}}

\begin{document}
\title{Recent Developments in Gravitational Microlensing}   
\author{Andrew Gould}   
\affil{Ohio State University}    

\begin{abstract} 
Twenty-one years after Bohdan's seminal paper
launched the field of gravitational microlensing, it
has radically diversified from a method narrowly
focused on finding dark matter to a very general
astronomical tool.  Microlensing has now
detected 12 planets, including several that
are inaccessible by other search methods.  It
has resolved the surfaces of distant stars,
served as a magnifying glass to take spectra of
extremely faint objects, and revealed a number
of surprising phenomena.  I take a sweeping
look at this remarkable technique, giving equal
weight to its successes and to the tensions that
are continuing to propel it forward.
\end{abstract}


\section{Introduction}   

While the idea of microlensing goes back to the famous \citet{einstein36}
paper in {\it Science}, and is worked out in even greater detail in Einstein's
notebooks from 1912 \citep{renn97}, Bohdan \citet{pac86} was the first
to recognize that with the arrival of modern CCDs and the high-speed
computing required to analyze them, microlensing's time had come.

The focus of Bohdan's original paper on this subject was dark matter,
and it prompted two major surveys toward the Large Magellanic Cloud (LMC),
which are reviewed by Charles Alcock in this volume.  But Bohdan
was always looking to push microlensing in new directions, 
most notably in his two
seminal papers that launched microlensing studies of Galactic
structure \citep{pac91} and microlensing planet searches \citep{mao91}.
Over the past 15 years, microlensing has developed as an important
tool in both these areas, and a third area as well: stellar atmospheres.

Parallel to this broad invasion of several areas of astrophysics,
microlensing activists pushed the field in a number of narrow,
rather arcane, directions, exploring weird higher-order effects such as
those due to finite source size, orbital parallax, terrestrial parallax,
xallarap, lens rotation, as well as degenerate solutions, and microlensed
variables.  One of the most exciting and unexpected developments
in microlensing has been that these weird effects, originally of interest
only to microlensing nerds, have started to interpenetrate with
the more mainstream investigations outlined in the previous paragraph.
This is because they provide additional information that is
of interest to a more general astronomical audience and make microlensing
applications more powerful.

In this contribution, I review some of these developments, pointing
to these interconnections whenever possible.

\section{Microlensing Basics}

\begin{figure}
\plottwo{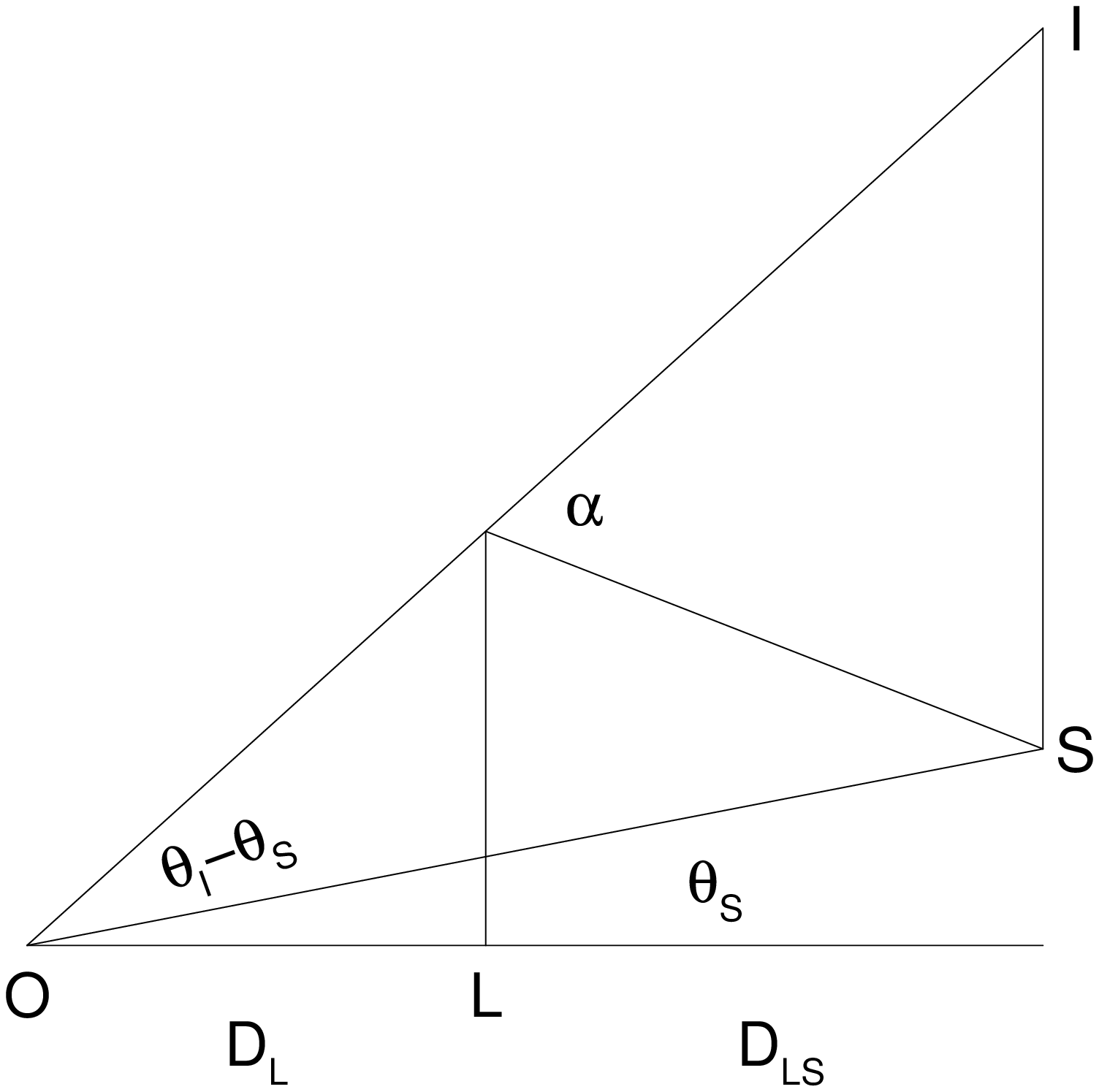}{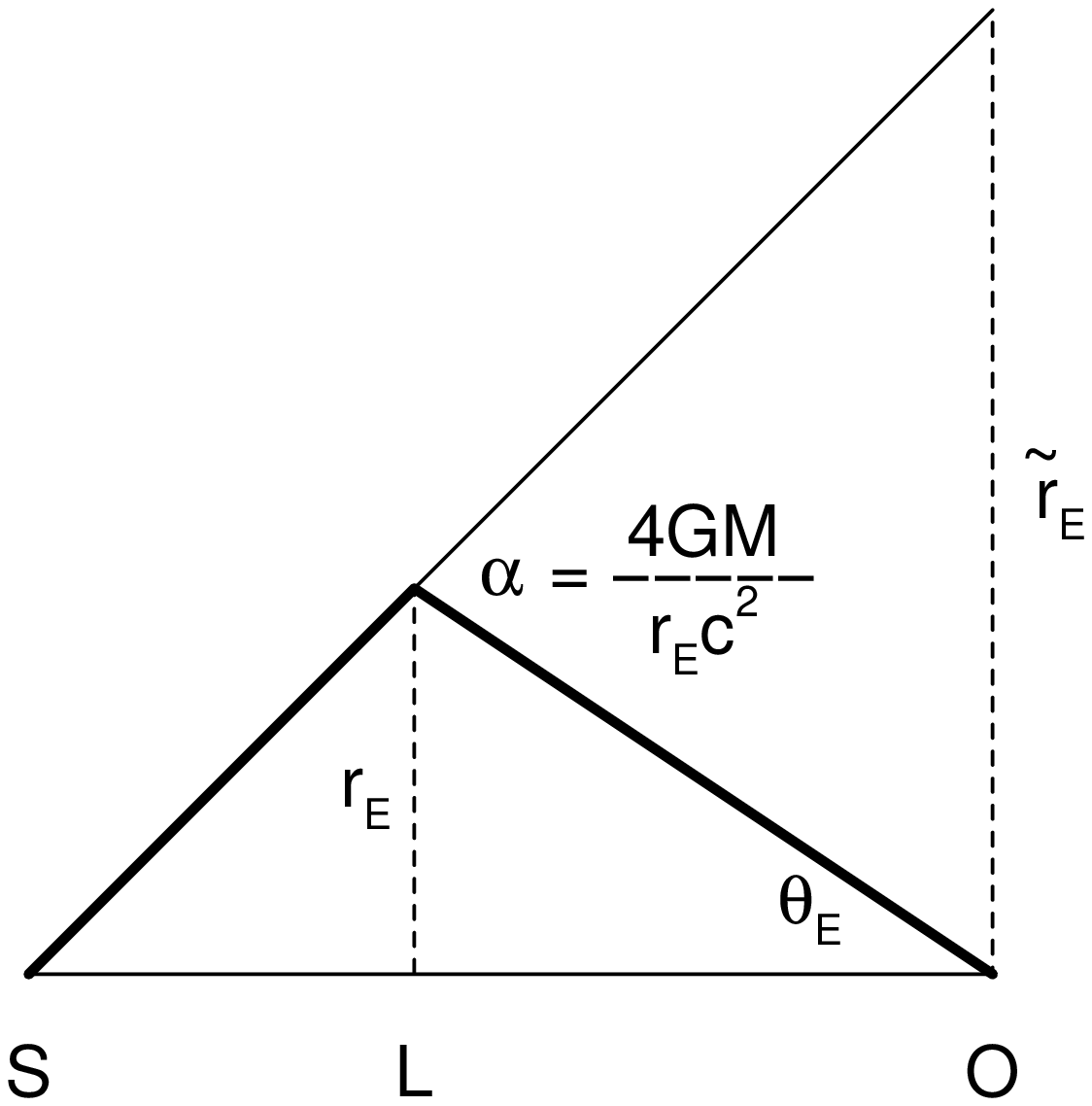}
\caption{\label{fig:lensdiags}
Left: Point-lens microlensing.  Mass (M) deflects light from source (S) 
by Einstein bending angle $\alpha= 4 GM/(c^2 D_{\rm L}\theta_{\rm I})$
to observer (O).  
Right: Relation of higher-order observables, the angular ($\theta_{\rm E}$) and
projected ($\tilde r_{\rm E}$) Einstein radii, to physical characteristics
of the lensing system.  Adapted from \citet{gould00}.
}
\end{figure}

It is a mark of the simplicity of point-lens microlensing that the 
basic results, including the main higher-order effects, 
can be encapsulated in two simple diagrams (Fig.~\ref{fig:lensdiags})
and a few simple equations.

Equating (from Fig.~\ref{fig:lensdiags}a)
$\overline{IS}=\alpha D_{\rm LS} = (\theta_{\rm I}-\theta_{\rm S})D_{\rm S}$,
where $D_{\rm S} = D_{\rm L} + D_{\rm LS}$, yields the quadratic equation,
$\theta_{\rm I}(\theta_{\rm I}-\theta_{\rm S})=\theta_{\rm E}^2$, which
sets the fundamental Einstein angular scale $\theta_{\rm E}^2\equiv
(4 GM/c^2)(D_{\rm L}^{-1}-D_{\rm S}^{-1})$.  The two solutions
are $u_\pm = (u\pm\sqrt{u^2+4})/2$, where 
$u\equiv \theta_{\rm S}/\theta_{\rm E}$ and
$u_\pm\equiv \theta_{\rm I\pm}/\theta_{\rm E}$ are scaled to $\theta_{\rm E}$.
Because surface brightness is conserved, the magnification $A$ is
given by the ratio of the combined area of the images to the area of the
source: 
\begin{equation}
A_\pm = \biggl|{u_\pm\over u}\,{du_\pm\over du}\biggr|,
\quad
A= A_+ + A_- = {u^2+2\over u\sqrt{u^2+4}}.
\label{eqn:lensmag}
\end{equation}

The two higher-order observables shown in Figure \ref{fig:lensdiags}b,
the angular ($\theta_{\rm E}$) and projected ($\tilde r_{\rm E}$) 
Einstein radii, can be measured if
the event can be compared to standard rulers on the sky and observer
planes, respectively. See \S~\ref{sec:masses}
These are then easily related to the
mass $M$ and the source-lens relative parallax 
$\pi_{\rm rel} = {\rm AU}(D_L^{-1}- D_S^{-1})$.  First,
$\alpha/\tilde r_{\rm E} = \theta_{\rm E}/r_{\rm E}$, so
$\theta_{\rm E}\tilde r_{\rm E} = \alpha r_{\rm E} = 4GM/c^2$.
Next, by the exterior angle theorem, 
$\theta_{\rm E} = 
\tilde r_{\rm E}/D_{\rm L} - \tilde r_{\rm E}/D_{\rm S}=
(\tilde r_{\rm E}/{\rm AU})\pi_{\rm rel}$.  In summary,
\begin{equation}
M={\theta_{\rm E}\over \kappa\pi_{\rm E}},\quad
\pi_{\rm rel}=\theta_{\rm E}\pi_{\rm E},\quad
\theta_{\rm E} = \sqrt{\kappa M\pi_{\rm rel}},\quad
\pi_{\rm E} = \sqrt{\pi_{\rm rel}\over\kappa M},
\label{eqn:mpirel}
\end{equation}
where $\kappa\equiv 4GM/(c^2{\rm AU})\sim 8.14\,{\rm mas}\,M_\odot^{-1}$
and $\pi_{\rm E}\equiv {\rm AU}/\tilde r_{\rm E}$.

\section{{Microlensing Planet Searches}
\label{sec:planets}}

\citet{mao91} showed that if a lens had a companion, it would distort
the primary lens's magnification field, inducing an ``astigmatism'' or 
``caustic structure'' near the peak.  These caustics are closed
contours of formally infinite magnification 
(see, e.g., Fig.~\ref{fig:ob05071}a, below): the magnification diverges
according to
a square-root singularity as the source approaches the caustic from the
inside.  The bigger the companion,
the bigger the caustic, and so the greater the chance that the
source would pass close enough to the lens
to be affected.  But their main
point was: even a planet could in principle be detected.

Of course, just as the planet perturbs the magnification pattern of
its host, the host also perturbs the planet field.  Since the host
is much bigger than the planet, this perturbation is also much bigger, so
a random source is much more likely to pass over the resulting
``planetary caustic'' than the ``central caustic'' highlighted by
\citet{mao91}.  This fact led \citet{gouldloeb92} to focus on
planetary caustics the next year when we advocated a search+followup
strategy for finding planets.  Microlensing events are extremely
rare (optical depth $\tau\sim 10^{-6}$), so huge areas must be
surveyed each night, which limits the number of observations of
each field.  But since the planetary perturbations are extremely
short $t_p\sim (M_{\rm planet}/M_{\rm Jupiter})^{1/2}\,$day,
the events that are found must be intensively monitored by other,
``followup'' telescopes scattered around the globe, in order to trace out
the planetary signature.

Although hardly noticed at the time, this subtle difference in
emphasis between these two papers grew into a major divergence, which
has since 
dominated all issues connected with microlensing planet searches.

\subsection{1st Microlensing Planet -- Pure-Survey Jupiter}

In 1995, Penny Sackett formed the PLANET collaboration \citep{albrow98}
to carry out this survey+followup strategy, but it was not until 2003 that the
first planet was discovered, OGLE-2003-BLG-235/MOA-2003-BLG-53Lb,
and this was by the survey teams themselves, not the followup groups
\citep{ob03235}.  Why?  The event had a 7-day 
planetary deviation, so the nightly survey data were basically adequate
to characterize it, which would not have been the case had it lasted
just 1 day (or less), as expected.  The perturbation was long because
the planet was sitting right next to the Einstein ring, and so induced a 
big caustic.  Such alignments are rare, but the survey groups are well
poised to find them because they monitor of order 600 events per year.
The followup groups, by contrast, monitor only the few dozen ``most
promising'' events.  The survey-group discovery of
the first microlensing planet was the first piece of evidence that
the survey+followup strategy originally advocated by \citet{gouldloeb92}
would require radical rethinking if it were to be successful.

\subsection{{High-Magnification Events}
\label{sec:highmag_events}}

In the meantime, \citet{jaro02} found a planet-candidate based on a single
deviant point, which consequently could not be confirmed.  This prompted
Andrzej Udalski (see these proceedings) to develop the OGLE 
``Early Early Warning System (EEWS)'', which would alert the OGLE observer
when an already-identified event was behaving ``unusually'', thereby
enabling OGLE both to alert the community and to carry 
out ``auto-followup'' observations itself.  This system actually went
off on OGLE-2004-BLG-343, a spectacular magnification $A=3000$ event,
but unfortunately the alarm was ignored by the observer.  However,
\citet{ob04343} showed that if this event had been properly monitored,
it would have had excellent sensitivity to Earth-mass planets, and
even some sensitivity to Mars-mass planets.  That is, the
``central caustic'' (i.e. high-magnification) events originally
highlighted by \citet{mao91} were actually much better targets
than the larger-caustic events singled out by \citet{gouldloeb92}.
Even though the caustics (and so the number of caustic-crossing
events) are smaller, the events in which this happens can be identified
{\it in advance}, enabling intensive followup right in the period
of greatest sensitivity.  Actually, this same point had previously been made
by several theorists \citep{griestsafi,rattenbury02}, but
as often happens, it was the practical demonstration that had the
biggest impact.

Another, completely unrelated development, pushed the Microlensing Follow Up
Network ($\mu$FUN) in the direction of high-mag events.
Jennie McCormick, a New Zealand amateur, sent me an email one day saying 
``I have data on your event, what do you want me to do with it?''
Of course, it seemed preposterous that a $12''$ telescope in one
of the wettest places in world could make a material contribution,
but I started sending her our microlensing alerts.  She contacted
Grant Christie, another NZ amateur, who ultimately made contact with
almost a dozen other amateurs around the southern hemisphere.
As these amateurs had to work during the day, we had to limit
requests to only the most sensitive events, generally high-mag
events.  Eventually, we realized that even at our professional-class
telescopes, we were wasting our time following non-high-mag events.
By 2005, our conversion was complete.

\subsection{2nd Microlensing Planet -- High-Mag Jupiter}

\begin{figure}
\plottwo{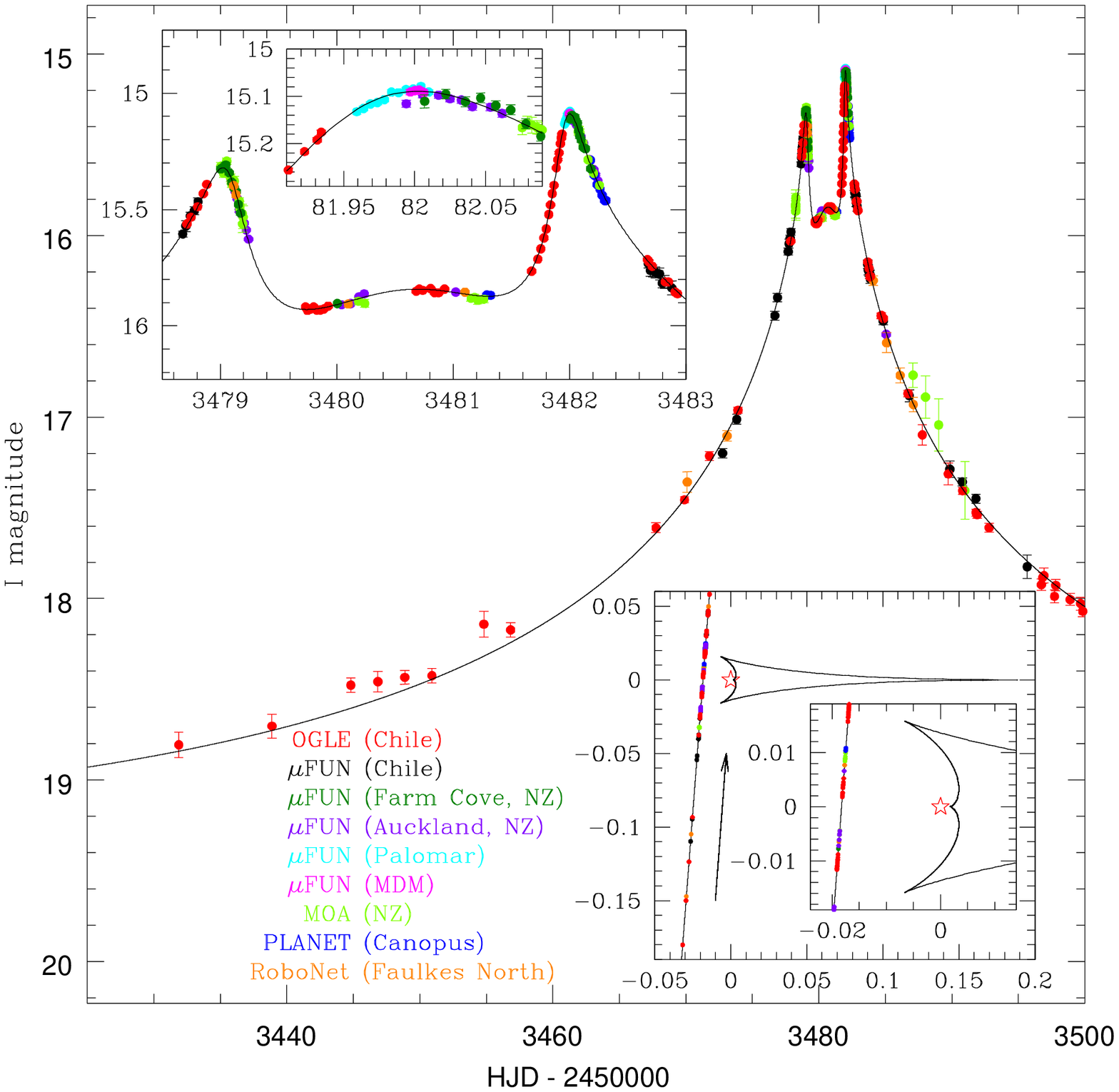}{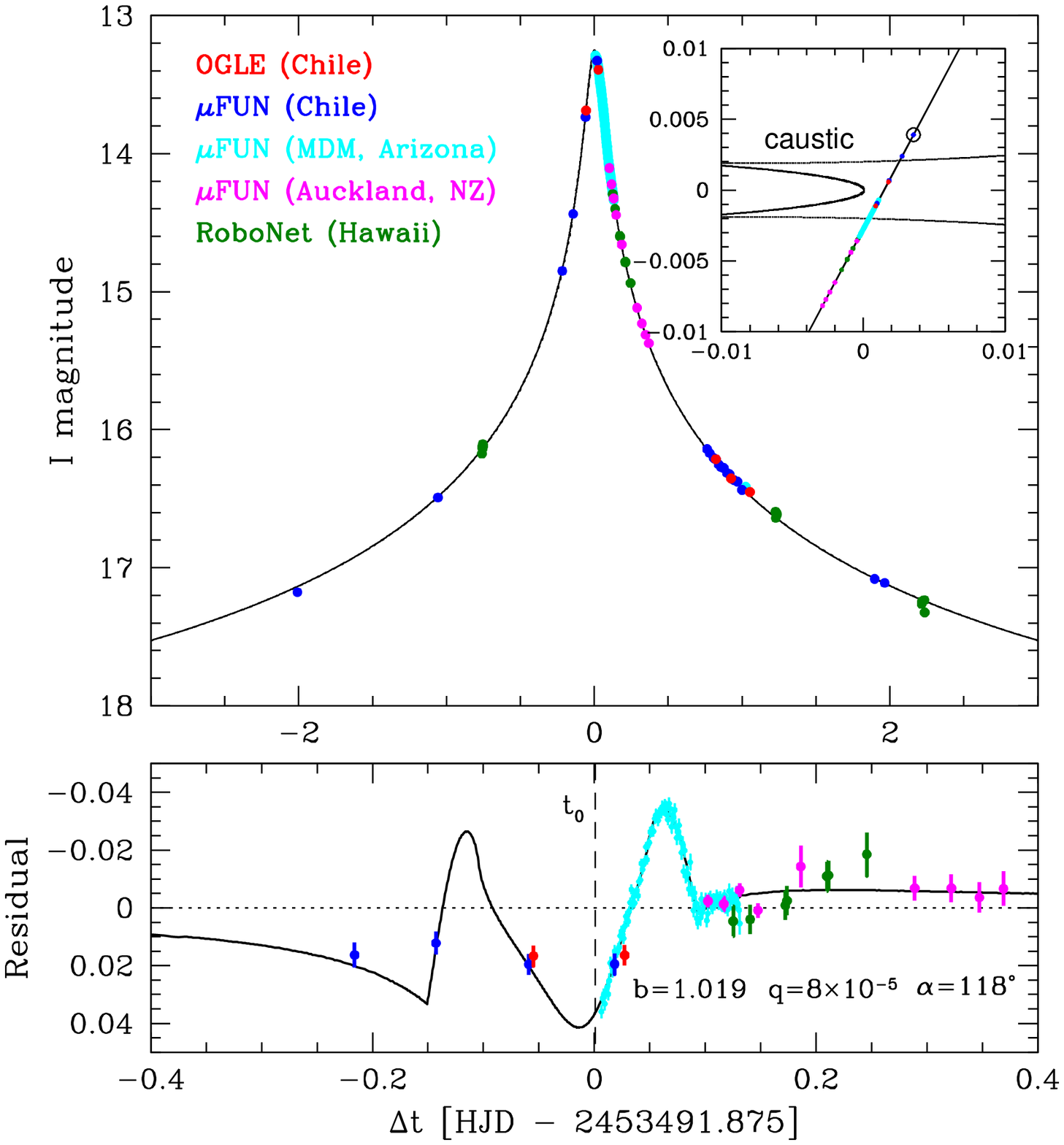}
\caption{\label{fig:ob05071}
Left: Jupiter-mass planet in high-mag event OGLE-2005-BLG-071.
Two major peaks and small peak in middle (upper inset) imply
source passes by two major cusps and a weak cusp in between
(lower inset).  This caustic geometry can only be produced by
planetary companions, in this case with mass ratio $q=7\times 10^{-3}$.
 From \citet{ob05071}.
Right: Neptune-mass planet in high-mag event OGLE-2005-BLG-169.
Upper panel shows ``basically normal'' event, but residuals
to point-lens fit reveal 2\% deviations.  Detailed modeling
is required to uncover the caustic structure (inset) due to
planet with $q=8\times 10^{-5}$, i.e. almost 100 times smaller
than OGLE-2005-BLG-071.  From \citet{ob05169}.
}
\end{figure}

The first fruit of this new strategy came early the next year when
OGLE-2005-BLG-071 started approaching high magnification.  Both
OGLE and $\mu$FUN Chile intensively observed the event as it approached
its peak until observations were cut off by dawn.
Shortly thereafter,
however, Jennie and Grant began observing on their $12''$ and $14''$
scopes (see Fig.~\ref{fig:ob05071}a).  
Over four nights, OGLE and $\mu$FUN telescopes traced out a triple-peak event:
two big peaks flanking a small peak in the middle, implying that the
source passed by a caustic with three cusps: strong, weak, strong
(see lower inset to Fig.~\ref{fig:ob05071}a).
It can be proved mathematically that such a geometry can only be
produced by a planet.  Jennie's comment: ``It just shows that you 
can be a mother, you can work full time, and you can still go out
there and find planets.''

\subsection{{3rd Microlensing Planet -- Survey+Followup Super-Earth}
\label{sec:ob05390}}

The PLANET collaboration has dedicated access to 4 1m-class telescopes
for May--August.  This caused them to
miss OGLE-2005-BLG-071, which peaked in April, but enables them
to follow many more events during the 4-month ``high season'',
i.e., not just the rich but rare high-mag events, but the run-of-the-mill
events originally advocated by Avi and me.  One of these,
OGLE-2005-BLG-390 showed a second bump well after peak.
The rounded shape of this bump implies that its full duration,
$2t_p\sim 0.6\,$days is dominated by the size of the source
rather than the caustic.  This is expected because the source
was very bright and red, hence very big.  Under these conditions,
it is straightforward to show that the planet/star mass ratio is approximately,
$q = (A_p/2)(t_p/t_{\rm E})^2$, where $A_p$ is the amplitude of the
second bump and $t_{\rm E}=10\,$days is the Einstein timescale.  That is,
one can simply read off the lightcurve, without any analysis, 
$q=9\times 10^{-5}$.  In fact, detailed analysis \citep{ob05390}
yields $q=8\times 10^{-5}$, corresponding to 5.5 Earth masses at the
estimated $M\sim 0.2\,M_\odot$ mass of the host.  This is also
the first event for which both survey and followup were absolutely required.
Both of the previous planets had perturbations lasting several days,
which allowed them to be basically characterized from survey data alone,
even though the followup data did substantially improve the characterization
in the case of OGLE-2005-BLG-071Lb.

\subsection{{4th Microlensing Planet -- High-Mag Neptune}
\label{sec:ob05169}}

Just a week after OGLE-2005-BLG-071 subsided, another event was
approaching peak, OGLE-2005-BLG-169.  In this case, OGLE did not observe
the event at all for 6 days before peak, the first 4 because of
weather and the last 2 because the telescope was dedicated to Chilean
observations.  Based on ``general suspicion'' that it might become 
high-mag, $\mu$FUN obtained some observations, but the night before peak,
the case was still not convincing:  $\mu$FUN (i.e., AG) failed to pursue
the event aggressively, but did ask Andrzej (who was at the OGLE telescope,
service observing for the Chileans) to sneak in an observation
of this event.  An email came back at 3 a.m.: the event was extremely
high-mag and there were no observations being taken!  I was asleep, but
heard the ``ping'' of my email and went upstairs to have a look.
I was quite dazed but eventually realized that the event could be
observed over peak from MDM, despite its northern location.  I called
up the observer who happened to be an OSU grad student, Deokkeun An.
I implored him to take time out of his own observing to obtain 9 images
of this event over the next 3 hours.  Recognizing that my request was much
too timid, Deokkeun actually took over 1000 observations, which traced
out a 2\% deviation from a magnification $A=800$ event 
(see Fig.~\ref{fig:ob05071}b).  As in a number of other microlensing events, 
the initiative of the observer proved crucial!
Exhaustive
analysis eventually demonstrated that this was a ``cold Neptune''
with $q=8\times 10^{-5}$.

\subsection{5th+6th Microlensing Planets -- Jupiter/Saturn System}

On 28 March 2006, the OGLE EEWS noted a tiny 0.1 mag deviation in
the previously unremarkable lightcurve of OGLE-2006-BLG-109, but
Andrzej was confident enough to issue a public announcement:
``Because short-lived, low amplitude anomalies can be a signature of a
planetary companion to the lensing star (cf. OGLE-2005-BLG-390)
follow-up observations of OGLE-2006-BLG-109 are strongly encouraged!!!''
This triggered observations from MDM only a few hours later, which
ultimately were important, but the event quickly returned to normal.
A few days later, however, it was clearly becoming high-mag, and so drew
many observations.  Grant Christie caught what seemed like a caustic
exit at magnification roughly $A=500$, 8 days after the first deviation,
which definitely raised the excitement level.  Within hours, Scott Gaudi
had a tentative model.  He drew a 6-sided (or 6-cusp)
caustic due to a Saturn-mass-ratio
planet.  The first small bump occurred when the source passed by a cusp.
Somehow the source had entered the caustic without being noticed and
had just exited.  Scott's trajectory would take the source by another cusp
3 days later, so he predicted another bump at that time.  However, reports
soon came in from the Wise observatory that the event was rising again,
and hours later
OGLE observations showed that it was again falling.  This new bump,
just 12 hours after Grant's ``caustic exit'', seemed to contradict
Scott's 3-day prediction.  Nevertheless, after 3 days, Scott's predicted bump
did occur: the Israel/Chile bump had been due to another planet, this one
of Jupiter mass-ratio.

\begin{figure}
\plotone{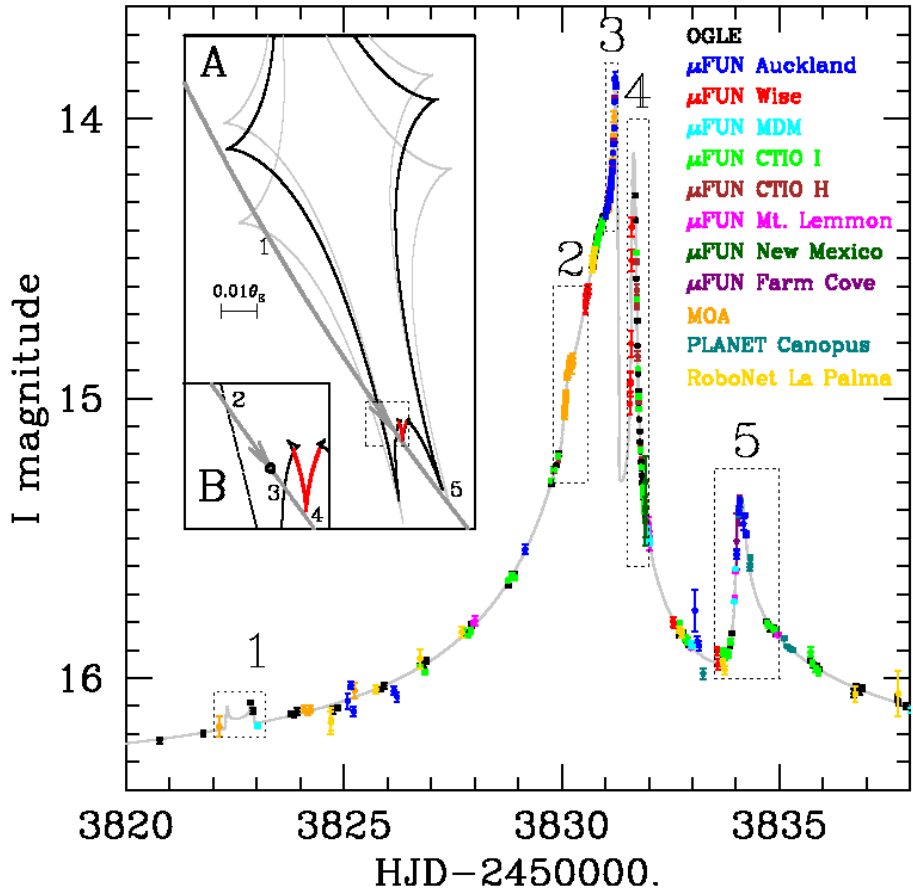}
\caption{\label{fig:ob06109}
First Jupiter/Saturn analog.  This spectacular lightcurve of OGLE-2006-BLG-109
has 5 distinct lightcurve features, which together reveal two planets.
Features 1, 2, 3, and 5 come from the black portion of the caustic
({\it inset A}) due to a Saturn mass-ratio planet very close to the
Einstein ring.  Feature 4, a sharp ``bump'' seen from Israel and Chile,
cannot be explained by this planet, but it
occurs very near the center of the lens geometry, just where perturbations
would be expected from other planets that are not near the Einstein ring
({\it inset B}).  This proves to have a Jupiter mass ratio. 
Because the Saturn is near the Einstein ring, its very small motion
leads to dramatic changes in the caustic between the time of Feature 1
({\it gray caustic}) and the time of Feature 3 ({\it black caustic}).
Twelve observatories contributed data, notably OGLE
(who announced Feature 1 in real time) and New Zealand
amateurs Grant Christie and Jennie McCormick, who caught the peak
at Feature 3. From \citet{gaudi08a}.}
\end{figure}

It took quite a while to fully decipher this event.  The Saturn
mass planet was very close to the Einstein ring.  In such cases,
the size of the caustic scales as $|b-1|^{-1}$, where $b$ is the
planet-star separation in units of the Einstein ring.  If $b\sim 1$,
then very small changes in $b$ can lead to large changes in the caustic.
Thus, the tiny planetary motion during the 8-day interval from the first
cusp approach to the caustic exit can lead to big changes in the caustic.
On the plus side, this means that if all these features can be 
properly modeled, one can measure some of 
the planet-orbit parameters, something no one thought would
be possible when microlensing planet searches were initiated.
On the minus side, analysis of the lightcurve requires very smart
algorithms applied to a supercomputer.
Dave Bennett took the lead in this analysis, eventually deriving
more comprehensive parameters for this system than any other.
It is a true Jupiter/Saturn analog, with similar mass ratios
and separation ratios as the solar-system gas giants.  The equilibrium
temperatures of these planets are also similar to Jupiter/Saturn, but
a bit cooler \citep{gaudi08a}.

It appears that in 2007, microlensers have discovered about 6 more planets,
but I have neither the space nor the permission to write about these
in detail.

\subsection{What Have We Learned About Planets?}

Given that microlensing has discovered only a handful of planets, 
compared to 250+ by other techniques, the
scientific payoff has been remarkably high.  The difference between
community expectations (which were rather dim) and what has actually
been achieved is due to two factors.  

\begin{figure}
\plottwo{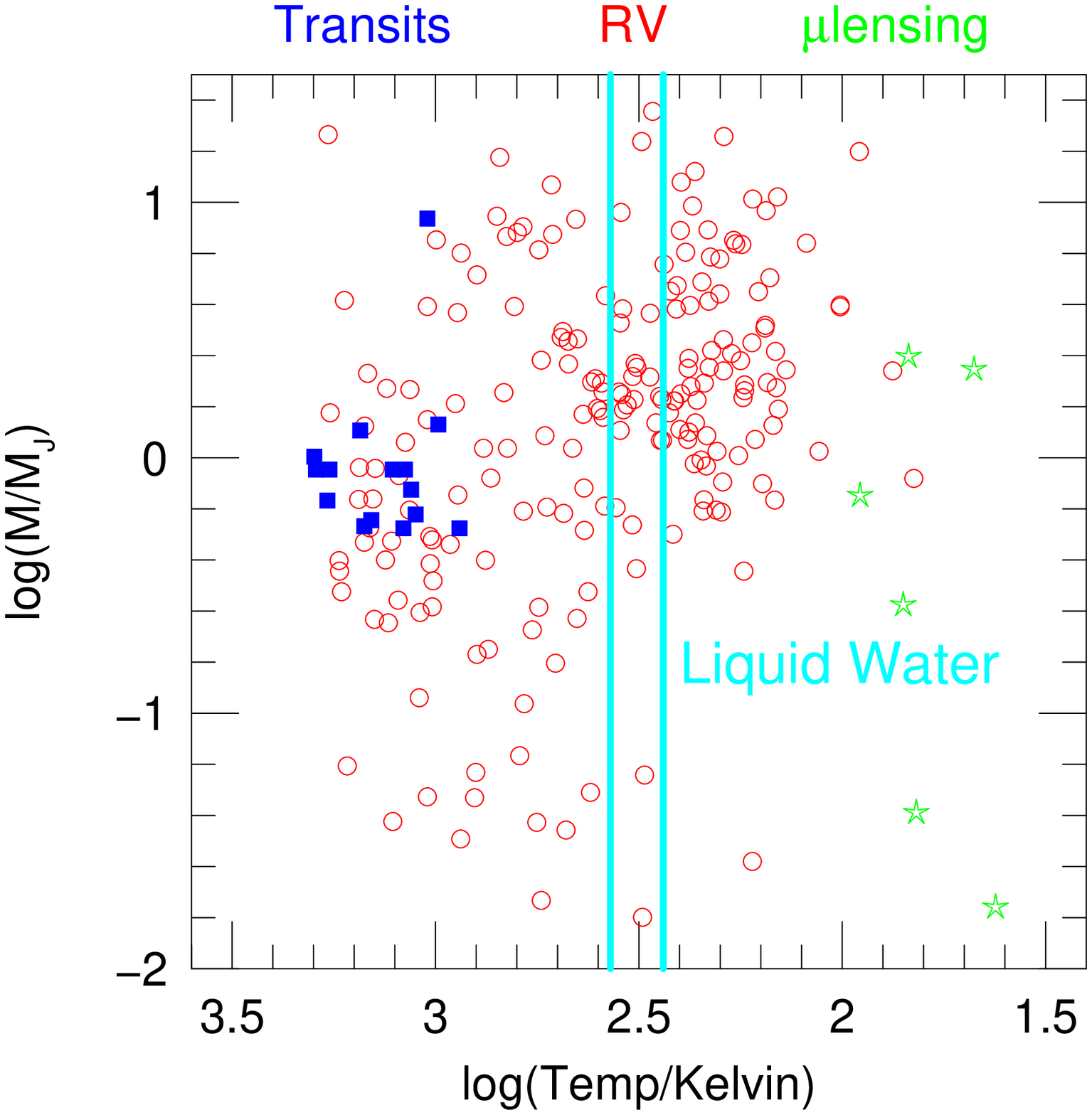}{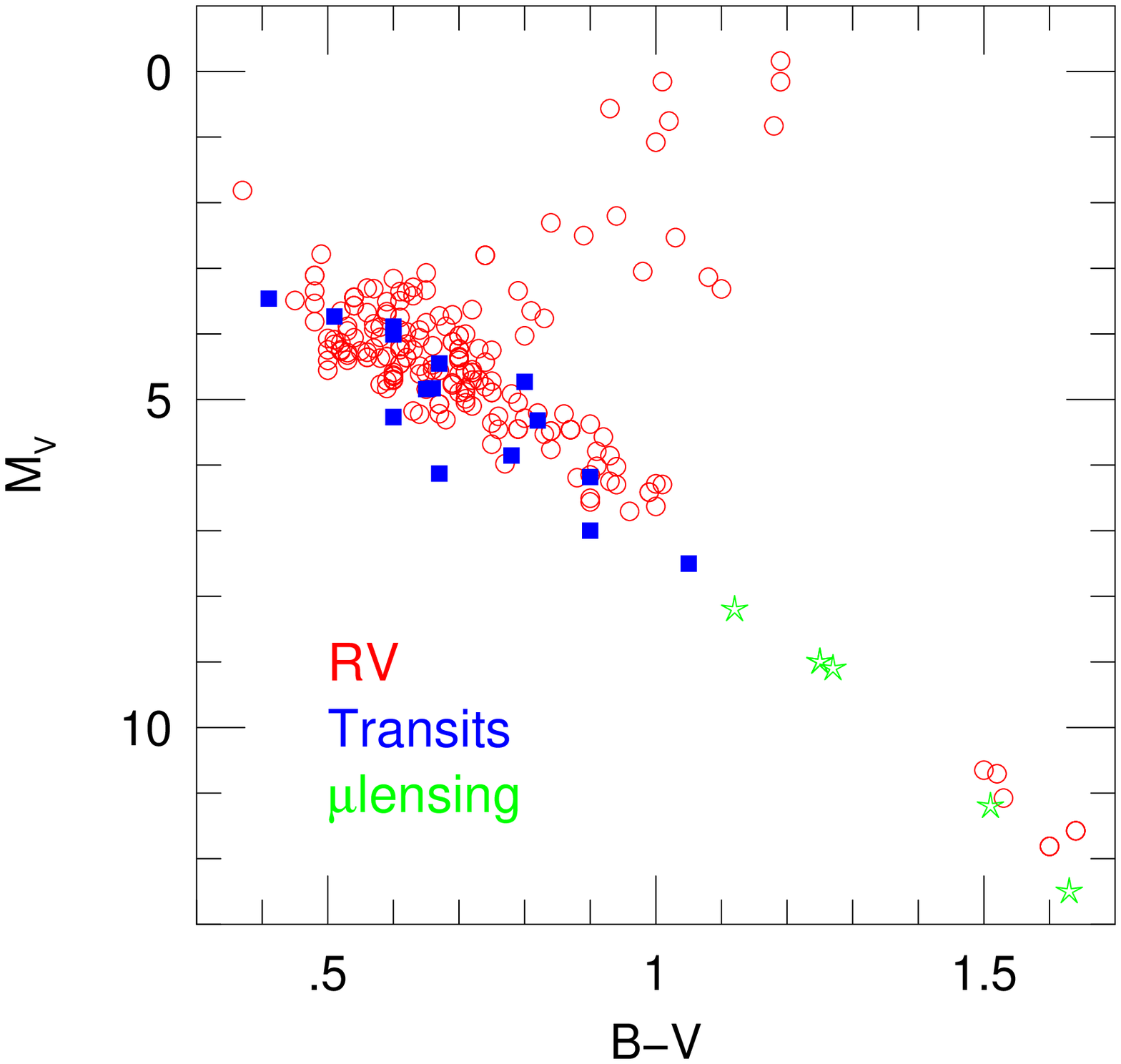}
\caption{\label{fig:cmd}
Left: Planet mass vs.\ equilibrium temperature of planets
detected by the Doppler ({\it red circles}), 
transit ({\it blue squares}), and microlensing ({\it green stars}) techniques,
as of June 2007.  Microlensing detects planets in the cold, outer regions
of their solar systems, where planet formation is expected to be most
robust.
Right: CMD of the host stars of microlensing planets.
Microlensing detects planets without serious
selection bias due to host properties.  It demonstrates that planet 
frequency in the outer regions is not strongly dependent on stellar type.}
\end{figure}

First, microlensing detections have yielded far
more information about the individual star-planet systems than had been 
thought possible.  Originally, it was believed that microlensing
detections would return exactly two pieces of information about
the system, the planet/star mass ratio $q$ and the planet-star projected
separation (in units of the Einstein radius $\theta_{\rm E}$) $b$.
Only the first quantity was regarded as truly interesting, since the
second could not be translated into a physical distance without knowing
both $\theta_{\rm E}$ and the distance to the lens $D_{\rm L}$.
In practice, we have generally been able to make pretty good estimates
of the host mass $M$ (and hence the planet mass $m=q M$), as 
well as $\theta_{\rm E}$ and $D_{\rm L}$ (and so the projected
separation $r_\perp=b D_{\rm L} \theta_{\rm E}$).  I will discuss
exactly how we do this in \S~\ref{sec:multiple}

Second, microlensing probes
a region of parameter space to which other methods are at present largely
insensitive, namely the cold regions out past the snow line,
where (at least according standard core-accretion theory) planet
formation should be most robust.  See Figure \ref{fig:cmd}a.
Microlensing is also essentially
unbiased by host mass, in sharp contrast to other methods.
Hence, as shown by the CMD of planet hosts (Fig.\ \ref{fig:cmd}b),
microlensing detects planets of the most common potential hosts, i.e.
late-type stars.

The fact that there are two microlensing detections of cold 
Neptunes/super-Earths
means that these planets are probably extremely common.  \citet{ob05169}
estimated that if all stars had planets in this mass range, and in
a 0.4 dex annulus bracketing the Einstein ring
of the host, then there would have been about 6 detections.  In fact
there were 2, indicating a rate of roughly 1/3 in this fairly narrow
range of radii.

Microlensing sensitivity scales roughly as planet mass.  There
are 4 Jovian-mass detections and two Neptune-mass detections,
and the two classes of planets differ in mass by about 1.5 dex
(see Fig.\ \ref{fig:cmd}a).  This indicates that gas giants are
of order 7 times less common than ice giants.

Of the 5 planetary hosts, one has two detected planets.  As discussed
above, these are close analogs of the Jupiter/Saturn pair that
dominate the mass in our own solar system.  Before planets were discovered,
it was generally believed that most solar systems would be like our own.
Then with the discovery of the pulsar planets and 51 Peg, weird planets
became more fashionable.  But the fact is, only microlensing actually
has sensitivity to Jupiter/Saturn analogs, so this is the only information
we have on how common they are.  Microlensing has detected Jovian-mass
planets around 3 stars (OGLE-2003-BLG-235/MOA-2003-BLG-53,
OGLE-2005-BLG-071, and OGLE-2006-BLG-109).  In the first of these,
there was a very low probability of detecting a second, Saturn-mass companion
had it been there.  In the second, there was a modest (roughly 30\%)
chance.  In the third there was an excellent chance (and it was actually
detected).  This  suggests that for systems where there is a Jupiter,
it may be highly likely that there is also a Saturn.

\section*{Higher-Order Microlensing Effects}   

Bohdan was fond of pointing out that microlensing is fundamentally
such a simple phenomenon that one could predict effects from first
principles and then go out and observe these effects in actual
events.  His favorite example of this was parallax.

\section{Microlens Parallaxes}

\begin{figure}
\plottwo{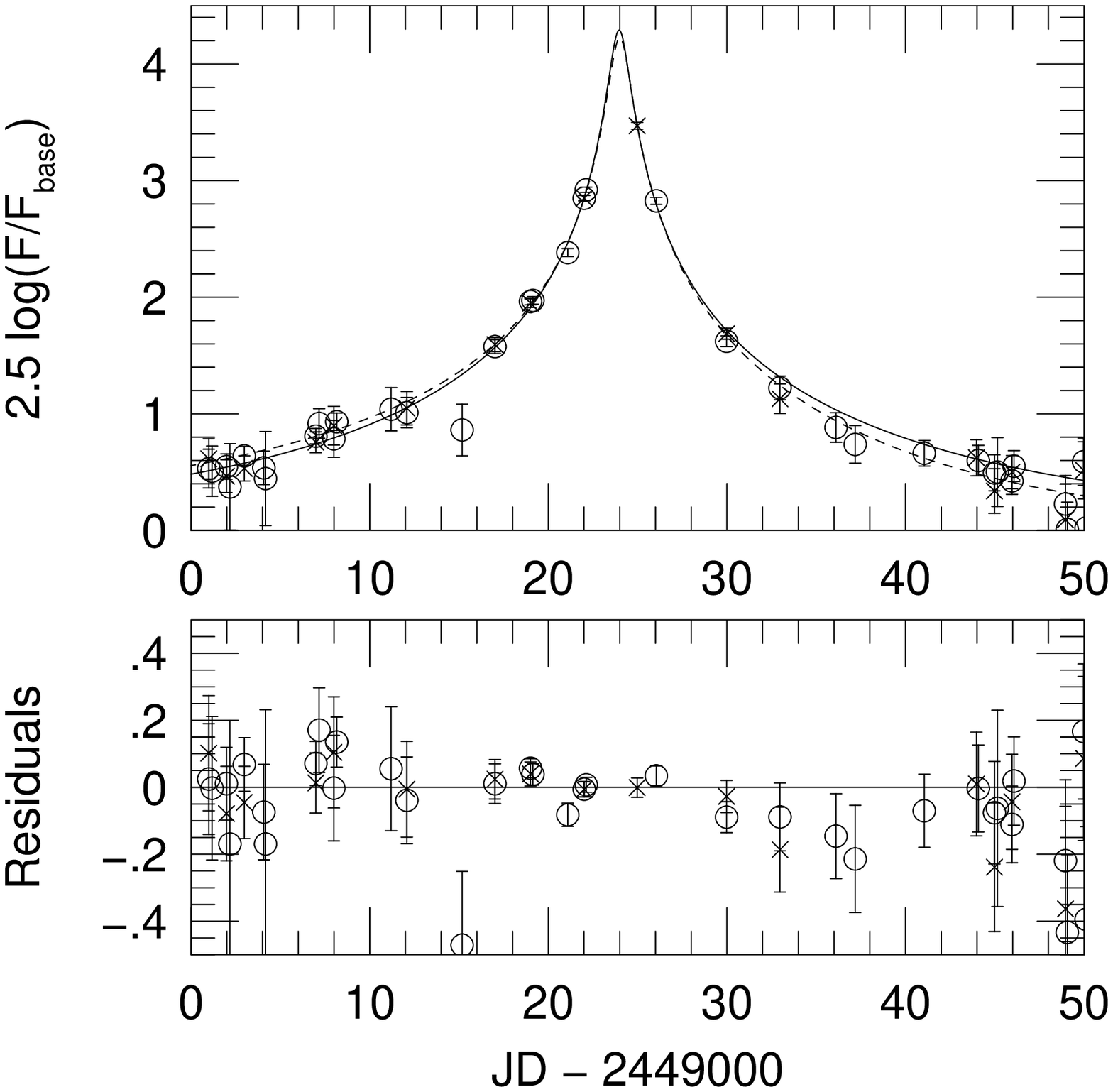}{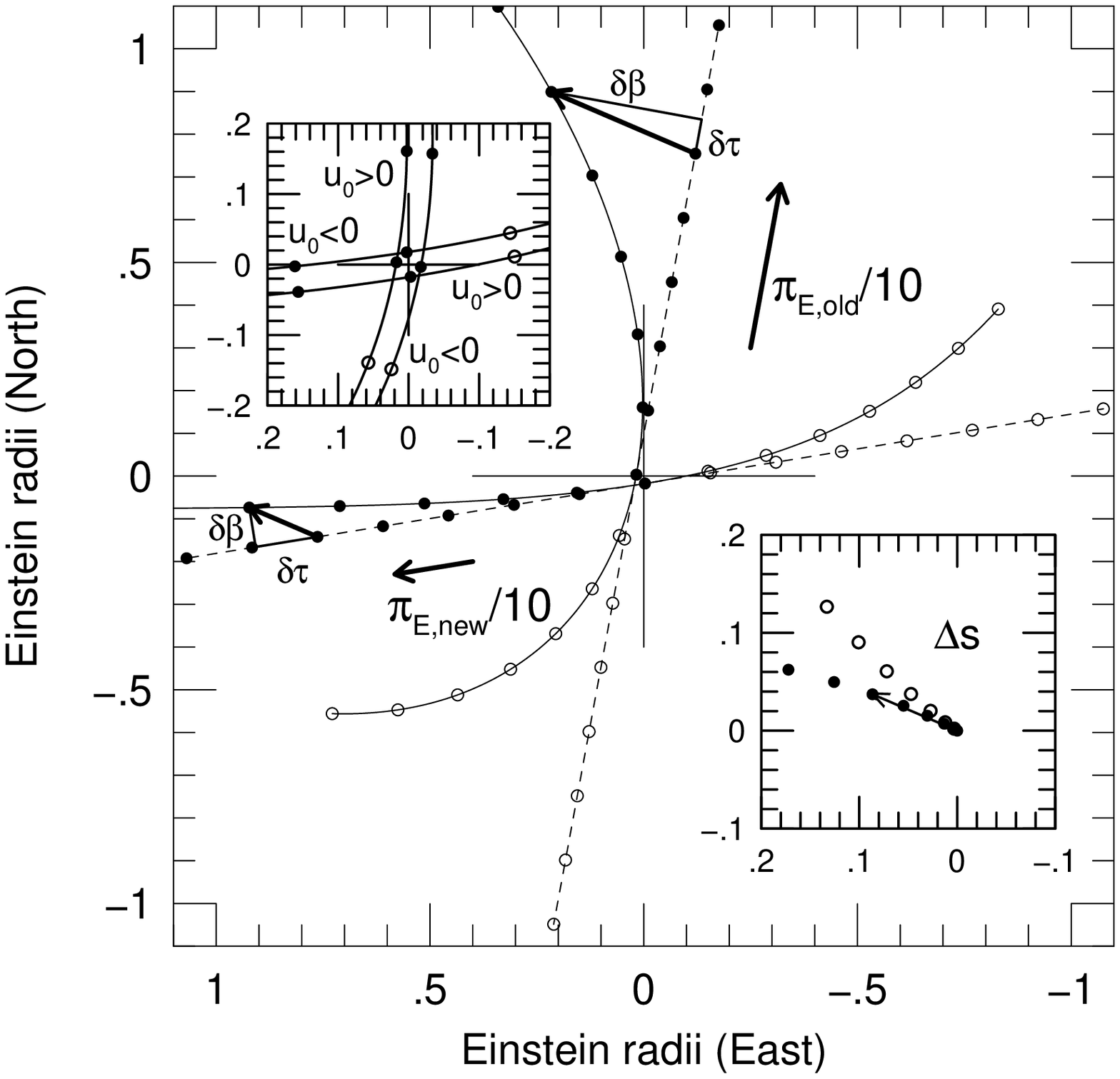}
\caption{\label{fig:lmc5}
Microlens parallaxes and degeneracies.
Left: Lightcurve of MACHO-LMC-5 shows clear asymmetry due to accelerated
motion of Earth, falling more rapidly than its rise.
Right: 4 possible trajectories of source-lens separation, all 
curved due to accelerated motion of Earth projected onto the
plane of the sky (lower inset). Deviations (from straight lines)
are proportional to $\Delta S$ (the accelerated displacement of the
Earth) and $\pi_{\rm E}$ (=AU/$\tilde r_{\rm E}$, the size of the
Earth's orbit relative to the projected Einstein radius).  The
fact that there are two sets of trajectories with radically different
directions and Einstein radii was totally unexpected, but is now
understood analytically.  From \citet{gould04}.}
\end{figure}

\citet{alcock95} made the first microlensing parallax detection
but parallax was actually observed in the very first microlensing event
observed toward the LMC, MACHO-LMC-5, although no one realized it
at the time \citep{alcock97}.  
Indeed no one realized it was the first event at the
time: hence its enumeration.  MACHO-LMC-5 was weird for other reasons:
the CMD position of the apparent source star does not coincide with
any LMC population.  \citet{gould97} had already suggested that this
``source'' was a foreground M dwarf and that it actually was the lens.  After
6 years, MACHO observed this and many other events with {\it HST}
and resolved two stars separated by about $0.1''$, a blue LMC star
that was clearly the source and a red foreground star, the putative
lens \citep{alcock00}.  But how could one be sure that the red star
was not just an unrelated foreground star?  

Dave Bennett went back to original
lightcurve and noticed the slight asymmetry (Fig.~\ref{fig:lmc5}a),
which led him to fit it for microlens parallax.  In analogy to
trigonometric parallax, the amplitude of the deviation from rectilinear
motion of the
source-lens trajectory in microlensing parallax is inversely proportional
to the size of what one is trying to measure, i.e., 
the projected Einstein radius $\tilde r_{\rm E}$.
So just
as one writes $\pi= {\rm AU}/D$ for trig parallax, it is convenient
to define the microlens parallax $\pi_{\rm E} = {\rm AU}/\tilde r_{\rm E}$.
However, in contrast to trig parallaxes, the microlens parallax simultaneously
measures the {\it direction} of lens-source relative motion.  So the
microlens parallax is actually a vector, $\bdv{\pi}_{\rm E}$.  Dave
found that the direction of $\bdv{\pi}_{\rm E}$ was the same as the
vector linking the red and blue stars \citep{alcock01}.  So the
red star {\it was} the lens, not just a chance interloper!  

This result had two important consequences.  First, it
showed that at least in this case, the lens was not part of a
putative dark-matter halo (``MACHO'') population, but was an ordinary
disk star.  Second, it led to the first mass measurement of an isolated
star, which I describe in the next section.

Microlens parallax measurements are relatively rare.  
\citet{poindexter05}
found only 22 events (out of about 3000 to that date) for which 
including parallax effects decreased $\chi^2$ by more than 100.
Nevertheless, microlens parallaxes have proved incredibly important,
as I discuss in \S~\ref{sec:planet_masses}

\section{{Microlens Masses}
\label{sec:masses}}

 From equation (\ref{eqn:mpirel}), one can determine the microlens
mass (as well as the lens-source relative parallax), if one can
just measure $\theta_{\rm E}$ and $\tilde r_{\rm E}$ (or $\pi_{\rm E}$).
As just mentioned, there are very few events for which $\pi_{\rm E}$
can be measured.  It also turns out that there are very few events for
which some ``angular ruler'' on the plane of the sky permits measurement
of $\theta_{\rm E}$.  The number for which the two measurements overlap is
minuscule.  Nevertheless, microlensing nerds have pursued microlens 
mass measurements like a holy grail, ultimately with major payoffs.  
The first microlens mass measurement was
EROS-2000-BLG-5 \citep{an02}, which I discuss in \S~\ref{sec:multiple}.

\subsection{{First Mass Measurement of an Isolated Star}
\label{sec:firstmass}}

The second was MACHO-LMC-5.  At one level this was trivial:
since \citet{alcock01} had measured the lens-source separation 
$\Delta\theta = 0.134''$ after $\Delta t=6.3\,$yrs, they could immediately
determine the lens-source relative proper motion 
$\mu_{\rm rel} = 21\,\rm mas\,yr^{-1}$.  Then from the measured Einstein
timescale $t_{\rm E} =21\,$ days, they could infer 
$\theta_{\rm E} = \mu_{\rm rel} t_{\rm E} =1.2\,$mas.  
Unfortunately, when combined with
their measurement $\pi_{\rm E}=4.2$ (and eq.~[\ref{eqn:mpirel}]), this gave
them a mass $M=0.036\,M_\odot$ and distance $D_{\rm L}= 200\,$pc, which
of course would be inconsistent with it being visible in the {\it HST}
image.

Bohdan played a significant role in the resolution of this puzzle.
\citet{smith03} developed an abstract formalism for analyzing
parallaxes, which (as referee) I found unexpectedly powerful.  They
Taylor-expanded the square of the source-lens separation to fourth
order in the approximation of uniform acceleration by the
Earth.  This led them to 
discover a degeneracy, which changed the trajectory from one side
of the Earth to the other (see Fig.~\ref{fig:lmc5}b, upper inset).  Because
this degeneracy basically left the magnitude of $\pi_{\rm E}$ unchanged,
it could not explain the ``wrong'' mass obtained by \citet{alcock01}.
However, by including jerk in the Taylor expansion, I discovered
a second, so-called ``jerk-parallax'' degeneracy, which did yield
a different $\pi_{\rm E}$ (see Fig.~\ref{fig:lmc5}b, main panel).
The distance implied by this solution $D_{\rm L}\sim 550\,$pc
was later confirmed by \citet{drake04} using trig parallax, and
the mass estimate $M=0.097\pm 0.016\,M_\odot$ is consistent with photometric
estimates \citep{gould04b}.

\subsection{{Multiple Paths to Microlens Mass Measurements}
\label{sec:multiple}}

There are basically 4 paths to the microlens parallax $\pi_{\rm E}$: 
Earth-orbital parallax, trigonometric parallax, Earth-satellite parallax,
and terrestrial parallax.  All four have been successfully employed.
There are also basically 4 angular rulers for measuring $\theta_{\rm E}$:
lens-source proper motion, finite-source effects, image resolution,
and centroid displacement.  The first of these is measured after the
event and the last three during the event.  Only the first two have
been successfully carried out.  As mentioned, microlens mass measurements 
require one from column A ($\pi_{\rm E}$) and one from 
column B ($\theta_{\rm E}$).

For MACHO-LMC-5, $\theta_{\rm E}$ was measured by lens-source proper motion,
but this is quite unusual: its proper motion was about 6 times larger
than typical lenses toward the bulge, there was a 6-year delay for the 
second epoch, and {\it HST} observations were still required to 
separately resolve the lens and source.
By far the most frequent measurement of $\theta_{\rm E}$ comes from
finite-source effects. If the source passes over a caustic, then
the lightcurve analysis automatically gives 
$\rho\equiv \theta_*/\theta_{\rm E}$, where $\theta_*$ is the angular
source radius.  The dereddened source color gives its surface brightness,
and the dereddened magnitude gives its flux, which together yield $\theta_*$.
Even in heavily reddened bulge fields, one can deredden the source
magnitudes by comparing the position of the source
to the clump on an instrumental
CMD (e.g., \citealt{ob03262}).  This was the method used in the
first microlens mass measurement, EROS-2000-BLG-5 \citep{an02},
which was a binary lens with an extremely well-covered caustic crossing.
Such caustic crossings are relatively rare, 
but what made EROS-2000-BLG-5 really unusual
was that it was also an extremely long event, which is what rendered 
it susceptible to Earth-orbital parallax.

The very largest $\theta_{\rm E}$ could in principle be resolved
using interferometry.  This would be a good way to confirm black-hole
candidates.  These have large Einstein rings, so are typically long
and so susceptible to Earth-orbital microlens parallaxes.  There is
an active program to do this at the VLT, but so far it has not been successful.
The fourth method will be described in \S~\ref{sec:sim}

 From equation (\ref{eqn:mpirel}), it is clear that if $\theta_{\rm E}$
is known, then the trig parallax directly yields 
$\pi_{\rm E} = \theta_{\rm E}/\pi_{\rm rel}$.  Hence, \citet{refsdal64}
already pointed out that microlens masses could be obtained from
trig parallaxes and proper motions.  More than 40 years later,
MACHO-LMC-5 is the only event to which this has been applied in practice
\citep{drake04,gould04b}.  Almost all remaining microlens parallaxes
come from lightcurve distortions (e.g., Fig.~\ref{fig:lmc5}a), but
there are two exceptions.

\begin{figure}
\plottwo{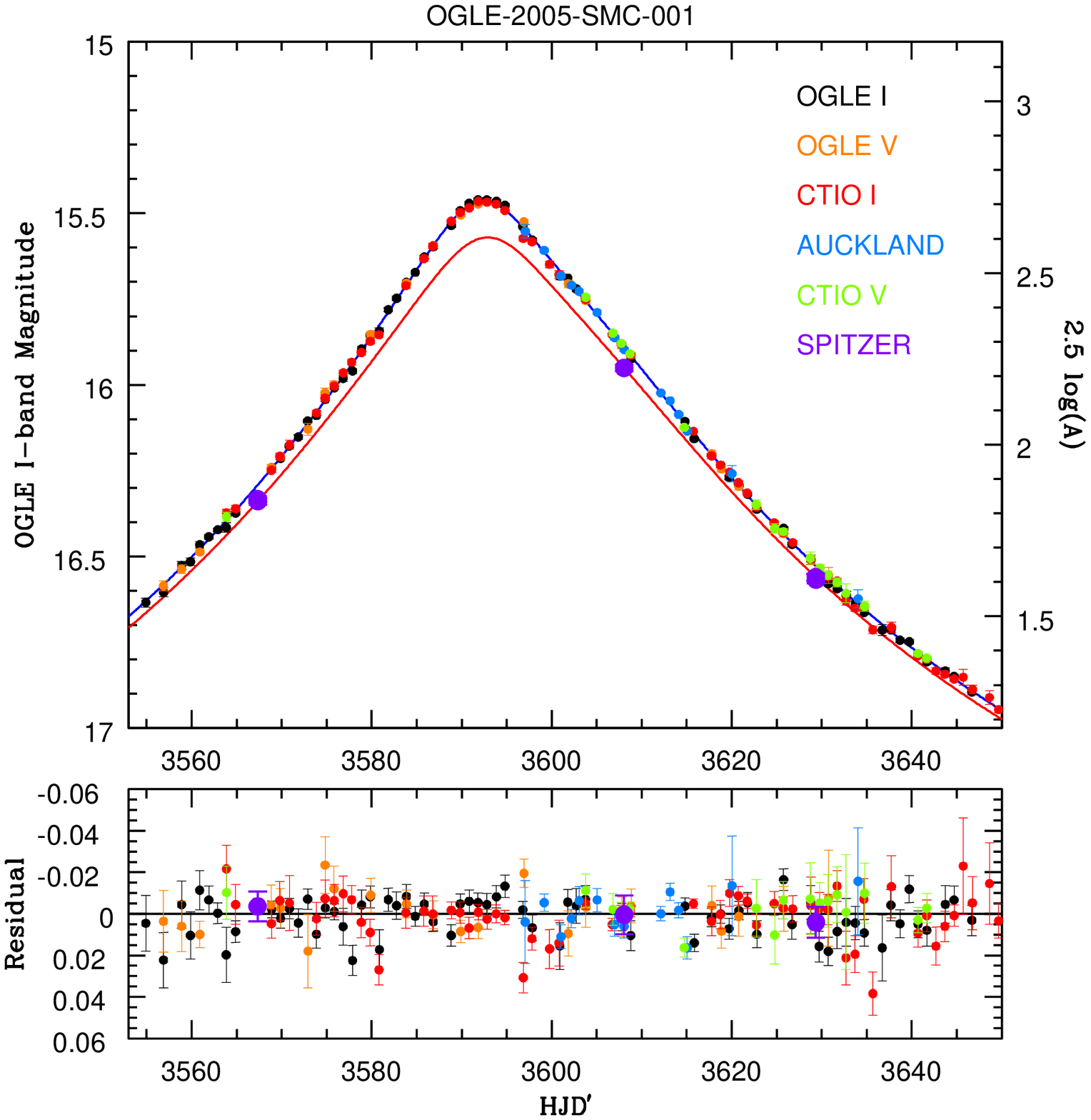}{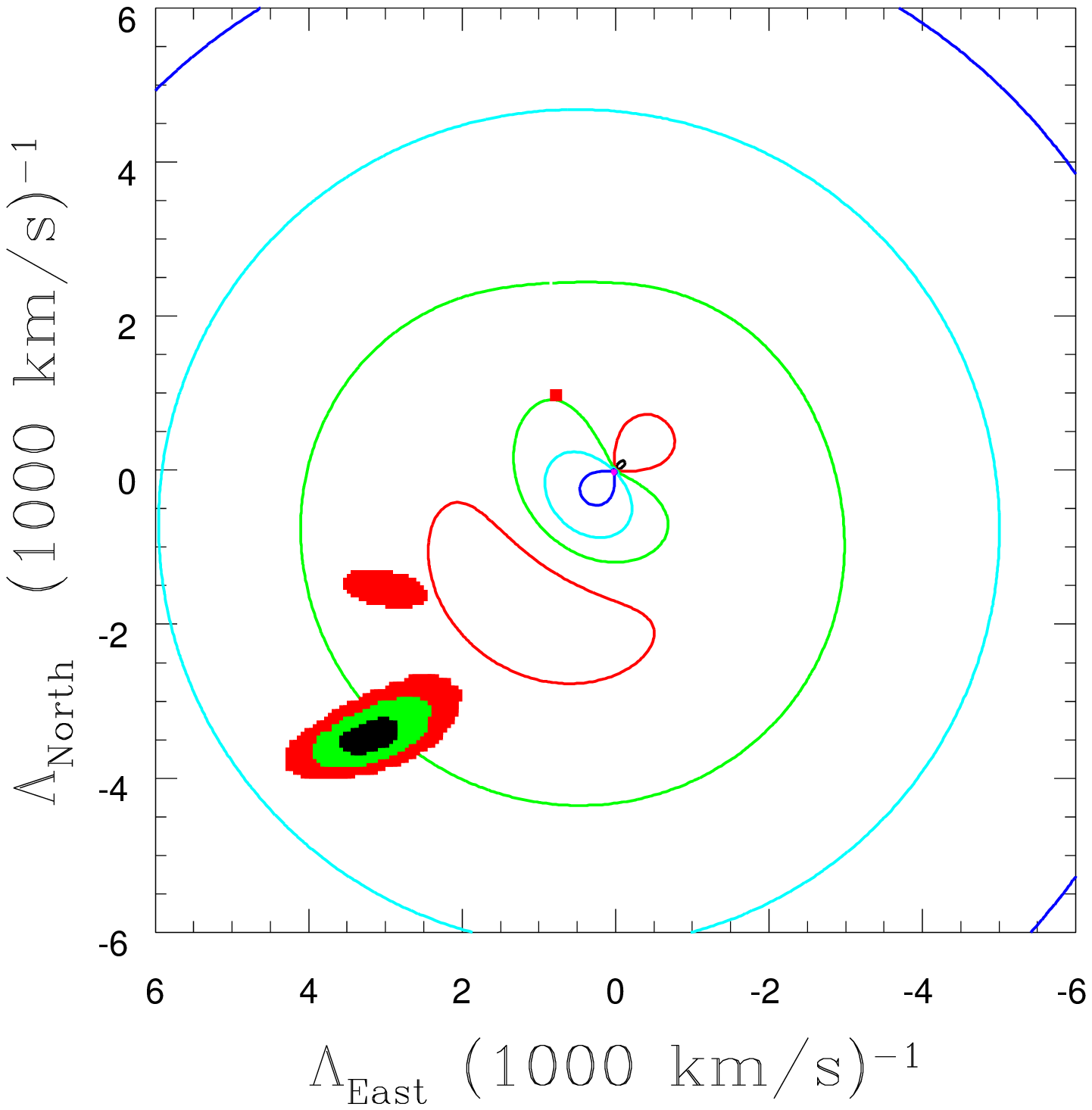}
\caption{\label{fig:oddpar}
First space-based parallax measurement.
Left: OGLE-2005-SMC-001 was observed by the {\it Spitzer} satellite
({\it lower curve, purple points}) when it was $\sim 0.25\,$AU from Earth.
Offsets in peak time (0.45 days) and peak flux (15\%) imply a
projected Einstein radius $\tilde r_{\rm E}\sim 30\,$AU.  
 Right: Inverse projected velocity 
$\bdv{\Lambda} \equiv \bdv{\pi}_{\rm E} t_{\rm E}/{\rm AU}$
of OGLE-2005-SMC-001 ({\it black, green, red} = 1, 2, $3\,\sigma$)
is near peak of likelihood contours ({\it red, green, cyan, blue} with
factor 5 steps) expected for halo lenses.  That is, the observed
$\tilde v \equiv \Lambda^{-1}\sim 230\,\rm km\,s^{-1}$ is close to typical
halo values ($\sim 450\,\rm km\,s^{-1}$), but an order of magnitude smaller
than typical SMC values (not shown). Adapted from \citet{dong07}.
}
\end{figure}

The very first idea for microlens parallax was to compare
lightcurves obtained from a satellite in solar orbit and the ground
\citep{refsdal66}.  Both the impact parameter and the time of maximum
would differ, enabling one to infer both components of $\bdv{{\pi}}_{\rm E}$.
\citet{dong07} made such a measurement using the {\it Spitzer} satellite
(see Fig.~\ref{fig:oddpar}), leading
them to conclude that the projected Einstein radius was very large,
$\tilde r_{\rm E}\sim 30\,$AU.  This could be either because the
lens was in the SMC or because it was a very massive ($10\,M_\odot$)
black-hole binary in the Galactic halo.  The latter was judged more
likely, since the projected velocity 
$\tilde v \equiv \tilde r_{\rm E}/t_{\rm E}\sim 230\,\rm km\,s^{-1}$
is typical for a halo lens but about an order of magnitude smaller
than expected for SMC lenses.  Still, since there was no measurement 
$\theta_{\rm E}$, this conclusion was not absolutely secure.

The reason ``solar orbit'' is usually required is that typically 
$\tilde r_{\rm E} \ga 1\,$AU, i.e., at least 20,000 times bigger than
the Earth.  This did not prevent two theorists,
\citet{holz96}, from pointing out that at least from the standpoint
of photon statistics, it would be possible to measure microlens parallaxes
from the lightcurve differences from two terrestrial observatories.
Amazingly, this has now actually been done as will be discussed elsewhere.

\subsection{{Future Routine Microlens Mass-Measurements with {\it SIM}}
\label{sec:sim}}

Given that it has proven so difficult to measure either $\pi_{\rm E}$ or
$\theta_{\rm E}$ separately, is it possible to {\it routinely} measure them
both together?  In fact, this would be possible with the 
{\it Space Interferometry Mission (SIM)} \citep{unwin07}.  
{\it SIM} would be in an Earth-trailing orbit, essentially the same
as {\it Spitzer}, and so could obtain Earth-satellite parallaxes in
exactly the same way.  But unlike {\it Spitzer}, it is capable of
routinely measuring $\theta_{\rm E}$ as well.   Although {\it SIM}
cannot generally resolve the separate images, it can measure the
astrometric displacement of the {\it centroid} of the images
relative to the source, 
\begin{equation}
\Delta\theta = 
{A_+ \theta_{\rm I+} + A_-\theta_{\rm I-}\over A_+ + A_-}
- \theta_{\rm S}
= {u\over u^2 + 2}\theta_{\rm E}.
\label{eqn:deltatheta}
\end{equation}
This reaches a maximum of $\theta_{\rm E}/\sqrt{8}$ at $u=\sqrt{2}$.
For typical bulge lenses, $\theta_{\rm E}\sim 300\,\mu$as,
so $\Delta\theta_{\rm max}\sim 100\,\mu$as, a very tiny angle.
But {\it SIM} precision is of order a few $\mu$as, meaning that
$\theta_{\rm E}$ (and so masses) could be measured to better than
10\%.  Thus, it would be possible to take a representive census of
all objects along the line of sight to the Galactic bulge, whether
dark (like black holes) or luminous (like stars).

\subsection{{Masses for Microlens Planets}
\label{sec:planet_masses}}

As should be clear from the last few sections, an enormous amount
of work has gone into developing and applying methods to find
$\theta_{\rm E}$ and $\pi_{\rm E}$, yet the few resulting mass measurements
have had little direct scientific payoff beyond proving to microlensing
nerds that we could do it.  In fact, however, there has been a big
practical payoff:  all 5 planetary hosts described in \S~\ref{sec:planets}
have masses and distances that are either measured or strongly constrained.
Hence, the 6 planet masses and projected separations are also measured
or well-constrained.
And it appears that this will also be true for the 6 microlensing planets
discovered in 2007. 
This seems initially implausible, since in general mass measurements
are so rare.  But first, planetary events are ``special'' in ways that
facilitate mass measurements.  And second, considerably more effort
(both observational and theoretical)
is expended by microlensers, once we know the event contains a planet.

\begin{figure}
\plottwo{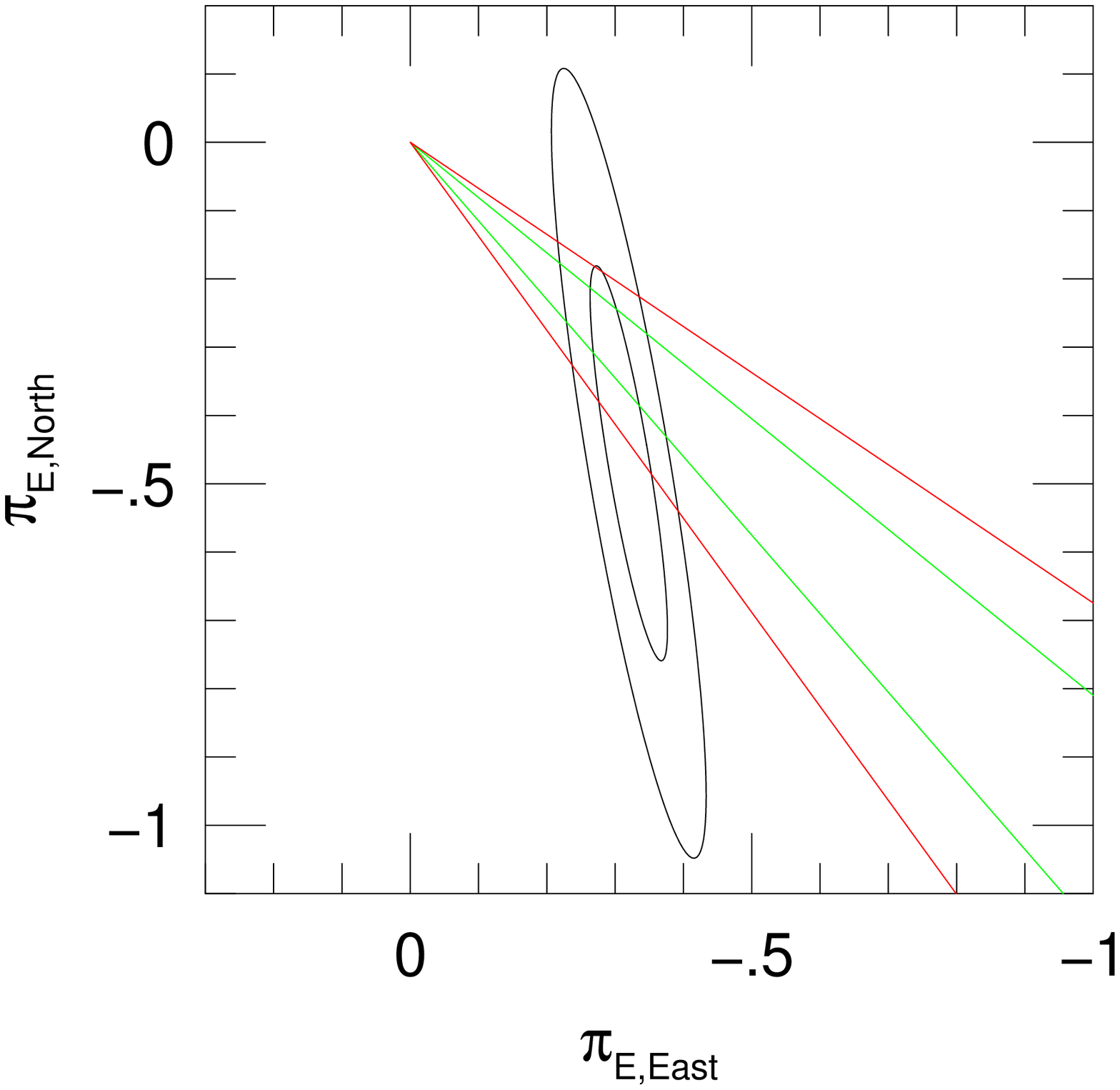}{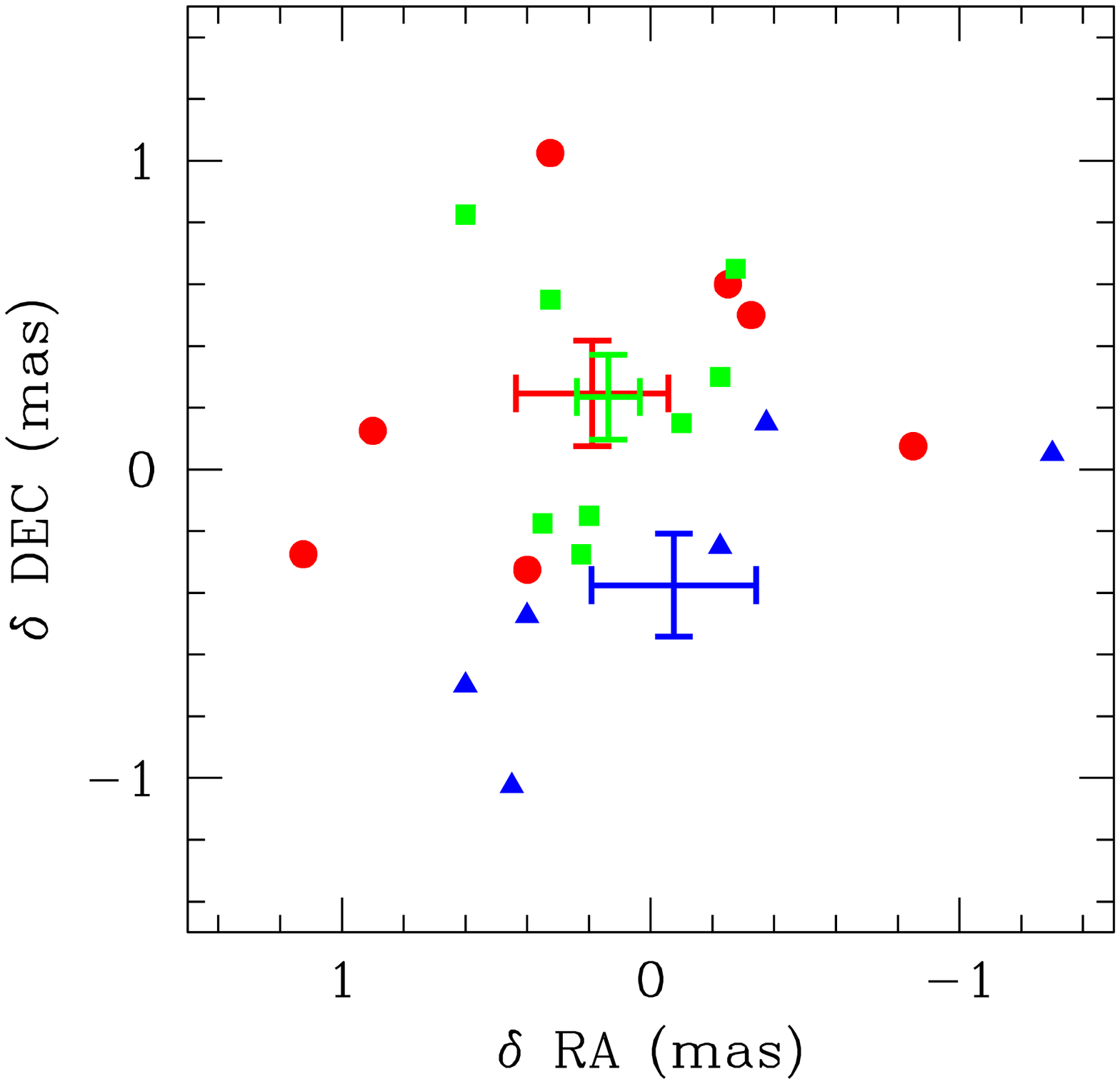}
\caption{\label{fig:ob03235}
Left: 1-$\sigma$ and 2-$\sigma$ contours for $\bdv{\pi}_{\rm E}$ ({\it black})
of OGLE-2005-BLG-071 (Fig.~\ref{fig:ob05071}a). Amplitude is
microlens parallax $\pi_{\rm E}$ and direction is that of lens-source
relative motion.  Only 1-D is well constrained.
Right: Relative source-lens centroids in 
$B$ ({\it blue}), $V$ ({\it green}), $I$ ({\it red})
of OGLE-2003-BLG-235/MOA-2003-BLG-53 from {\it HST} images
1.78 years after peak.  These yield direction of proper motion
($20^\circ$ north through east).  [In a perfect world (without errors) 
the $V$ point would be exactly aligned with the
axis connecting the $B$ and $I$ points, lying slightly closer to the
former.]
Left (again): A direction measurement from similar
observations of OGLE-2005-BLG-071 would resolve its 1-D parallax degeneracy 
(1-$\sigma$ [{\it green}] and 2-$\sigma$ [{\it red}] lines in left panel).
}
\end{figure}

The first point is that in sharp contrast to ``regular'' microlensing
events, $\theta_{\rm E}$ is almost routinely measured in planetary
events.  In ordinary events, the probability that the lens will
transit the source (thereby giving rise to measurable finite-source
effects) is $\rho=\theta_*/\theta_{\rm E}$, which is about $\rho\sim 1/500$
for main-sequence sources and typical lenses.  But in planetary events,
there is hardly any perturbation at all unless the source passes
very close to, or right over a caustic, so finite-source effects are
almost automatic.  To date, all planetary events have them, and they
are pronounced in all but OGLE-2005-BLG-071.

Second, due to another selection effect that could hardly have been
anticipated, a remarkably high fraction of planetary events have
detectable microlens parallax: 2 out of 5.  Four of the 5 events
were high-mag (due to selection effects described in 
\S~\ref{sec:highmag_events}).
These 4 have systematically longer timescales than typical OGLE events.
Specifically, they are longer than 79\%, 90\%, 95\%, and 97\% of them,
respectively.  And long
events display measurable parallaxes much more often, simply because
the Earth's motion deviates from a straight line during the event
as the square of the Einstein timescale.  Why are planets found
preferentially in long events?  One definite reason is that,
by definition, long events unfold more slowly, which increases the
chance that their high-mag character will be recognized in time
to initiate the dense monitoring required to find planets.
A second possible reason is that the planets we are finding seem to
be orbiting foreground disk stars, rather than bulge stars, and disk-lens
events tend to be longer than bulge-lens events.  At this point, we
cannot tell which way this selection pressure is working 
(i.e., if most planets are in the disk, which would bias us toward the
intrinsically longer disk events, or if the long events being better observed
is biasing us toward monitoring events caused by disk stars) or even if
the effect is real, but it is a possibility to keep in mind.

Whatever the exact cause, two events (OGLE-2006-BLG-109 and OGLE-2005-BLG-071)
have measurable parallaxes.  The first of these is quite good,
but the second is rather crude: its $\bdv{\pi}_{\rm E}$ 1-$\sigma$
error-ellipse axes are $(0.6\times 0.06)$.  See Figure \ref{fig:ob03235}a.
Under normal circumstances, 
we would not call this a ``measurement'' at all, or rather, we would
call it a ``one-dimensional parallax'' and move on.  The origin
of such 1-D parallaxes is easily seen from Figure \ref{fig:ob05071}a:
the lightcurve is asymmetric, being above the model on the rise and
below it on the fall.  This effect is caused by the uniform component 
of the Earth's acceleration toward the projected position of the Sun 
at the peak of the event,
and so very well constrains $\pi_{\rm E,\parallel}$, the component of
$\bdv{\pi}_{\rm E}$ in this direction.  Only longer, or much more precisely
photometered, events reveal the much subtler effects from motion
in the direction perpendicular to the Sun.

While the OGLE-2005-BLG-071 parallax ``information'' is rather
ambiguous on its own, it could be transformed in into
a genuine microlens parallax if, somehow, the direction of lens-source
motion (and so of $\bdv{\pi}_{\rm E}$) could be independently
determined.  And
this brings us to yet another type of information that is not usually
available for normal events: high-resolution post-event imaging.
We already saw that the source and lens were separately resolved
6 years after peak for MACHO-LMC-5 (\S~\ref{sec:firstmass}).  That
was only possible because the relative proper motion was exceptionally
fast and the observers were ready to wait an exceptionally long time.
But much smaller source-lens relative displacements can be detected
by taking advantage of the fact that the source and lens generally have
different colors.  This means that the {\it centroids} of $B$  and $I$
light will be displaced from one another as the source and lens separate, long
before the two stars are separately resolved.  The amplitude of the centroid
displacement is the product of the lens-source displacement and a
function of the $B$ and $I$ mags of the two stars.
Figure \ref{fig:ob03235}b
shows an example of this using post-event {\it HST} images of
OGLE-2003-BLG-235/MOA-2003-BLG-53.  In this case, the source-lens
displacement was known (from finite-source effects), so the
centroid offset yielded the color-function, which was used
(together with stellar color-mag relations and the source
flux as determined from the microlensing fit) to estimate the
lens mass.  However, similar measurements made several years
after the peak of OGLE-2005-BLG-071 could be applied to
reverse effect.  That is, to the extent that the color-function is
known (which it approximately is in this case from {\it HST} images
during the event), the centroid displacement would
give the lens-source displacement (and so $\mu_{\rm rel}$ and
hence $\theta_{\rm E} = \mu_{\rm rel} t_{\rm E}$).  And, more
importantly in the present context, {\it whether or not the color function
is known}, the centroid color displacement gives the {\it direction}
of $\bdv{\mu}_{\rm rel}$, which is the same as the direction of
$\bdv{\pi}_{\rm E}$.  Figure \ref{fig:ob03235}a shows the result
of a hypothetical future centroid-offset measurement for
OGLE-2005-BLG-071.  This offset ``picks out'' a narrow subset
of the parallax solutions, transforming the 1-D parallax derived
from the lightcurve into a 2-D parallax.

While no such late-time astrometry of OGLE-2005-BLG-071 has yet been 
obtained, Subo Dong (in preparation)
has made an incredibly detailed investigation of a variety of higher-order
effects, including the 1-D parallax just described, constraints on the 
proper motion from finite-source effects and from event and post-event 
{\it HST} imaging, limits on the presence of third bodies from the
lack of lightcurve distortions, and others.  Together these imply
that the lens star is probably a high-velocity, low-metallicity
(i.e., thick disk) M dwarf, which would be rather unexpected for
a planetary host.  Late-time astrometry would confirm (or contradict)
these tentative conclusions.

In brief, by taking advantage of intensive observations during the
event, taking carefully chosen high-resolution images during and
after the event, combining all available data, and performing
systematic cross-checks among them, it is often possible to
measure masses and distances accurate to 20\% or better.
Even when the only pieces of information are $\theta_{\rm E}$
and $t_{\rm E}$, it is still possible to combine these with 
priors for the lens and source distances and velocities to
derive a statistical estimate of the lens mass.  This was
the approach taken for the cold super-Earth
OGLE-2005-BLG-390 (\S~\ref{sec:ob05390}) and the cold Neptune
and OGLE-2005-BLG-169 (\S~\ref{sec:ob05169}).

\section{Binary Lens Revolution}

\begin{figure}
\plottwo{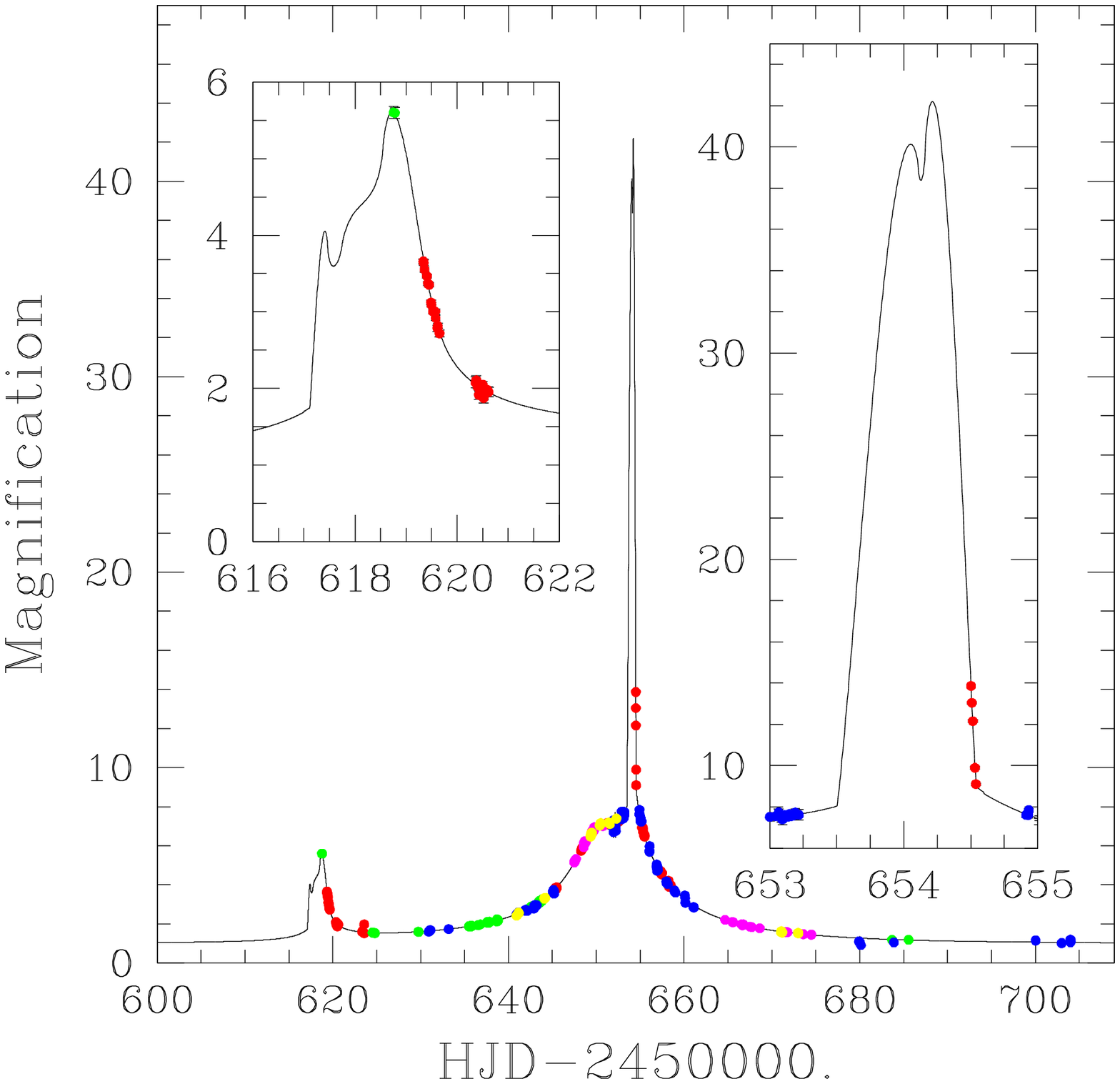}{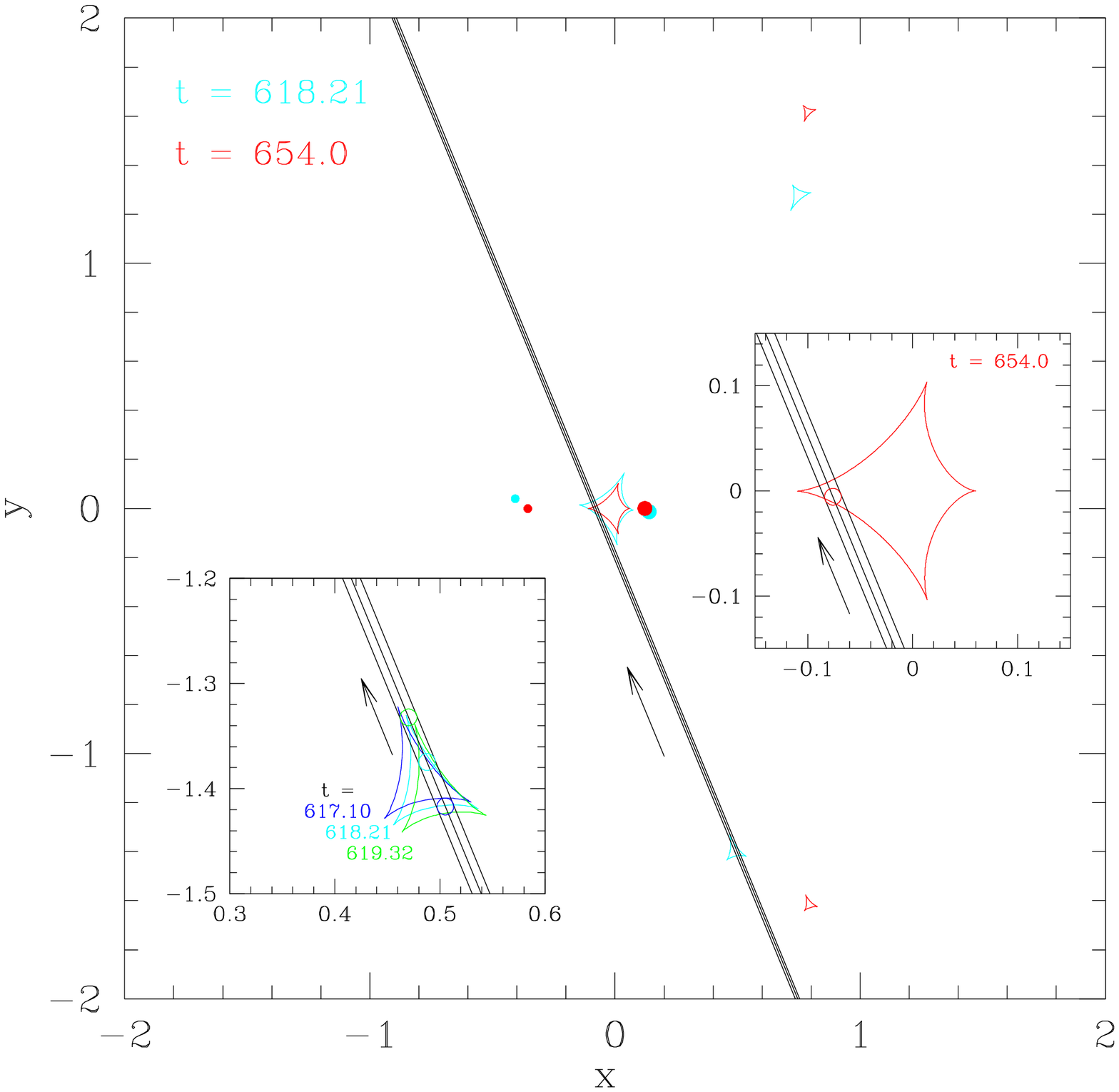}
\caption{\label{fig:mb9741}
Binary revolution in MACHO-97-41.
Left:  PLANET lightcurve and model \citep{albrow00}.  Two bumps
are due to outlying and central caustics of close binary, respectively.
``Wild'' features of model without data points are confirmed by
independent data set \citep{bennett99} [not shown].
Right:  In static model ({\it red caustics}) trajectory determined
from central caustic does not pass through outlying caustic.
Rotating model ({\it cyan caustics}) rotates and moves inward
the position of the outlying caustic so that it matches data.
}
\end{figure}

MACHO-97-41 was a curious event.  It showed a short, 3-day bump and
then seemed to return to normal.  A week later it began rising
sharply, briefly spiking to magnification $A=40$  before returning
to baseline. See Figure \ref{fig:mb9741}a.  The second bump is
easily fit as the central caustic of a close-binary lens (i.e.,
with both components inside the Einstein ring).  Such close binaries
always have two small outlying caustics in addition to the central
caustic, which would seem to explain the first bump.  The trouble
was, the predicted position of this small caustic had the wrong
distance from the central caustic to account for the timing
of the first bump and the wrong angular position to be
intersected by the source trajectory 
({\it red} caustic in Fig.~\ref{fig:mb9741}b)  
However, both effects are easily explained by ``binary revolution''.
To the extent that components move farther apart, the outlying caustic
will move closer to the central caustic.  And to the extent that they
rotate on the sky, the angular position of the caustic will change
({\it cyan} caustic in Fig.~\ref{fig:mb9741}b).  Hence, the very
peculiar geometry of this event permits measurement of the two 
instantaneous components of the binary internal motion on the plane
of the sky.

This example immediately raises two important questions.
First, how do we know that this very complicated model, which
predicts the incredibly elaborate (and largely unobserved)
lightcurve seen in Figure \ref{fig:mb9741}a, is actually correct?
And second, who (besides microlensing nerds) cares?

As it happens, in this case, two completely independent groups observed
this event and the model shown (based only on the data shown)
predicts the second data set almost perfectly.  This is a pretty
important test of the robustness of microlens models of complex
lightcurves with higher order effects.

As to the second question, at least for 10 years the answer was ``no one''.

\subsection{Planetary Lens Revolution}

Despite early predictions that a Jupiter-mass planet would generate
1-day deviations, 3 of the 5 planetary events have had 3--12 day 
perturbations.  The fundamental reason for this is that
the probability of detecting a planet is proportional to the size of
the caustic, so the relatively rare planets that are close to the
Einstein ring (and so have large caustics $\propto 1/|b-1|$) have
enhanced probability of detection and also proportionately longer
planetary perturbations.

The durations of these perturbations are still very short compared to
the typical orbital periods (several years), so one would not
at first sight expect any noticeable change in the caustic during
the perturbation.  But the very fact that the caustic size is
$\propto 1/|b-1|$, means that for $b\sim 1$,
small changes in $b$ lead to large
changes in caustic size.  Similar leverage applies to angular motions.
In the case of OGLE-2006-BLG-109, $b=1.04$.  Hence, a change
in $b$ of less than 0.5\% during the 8-day interval between the first
cusp crossing and the peak, would lead to a 10\% change in the
caustic size.  Far from being almost too subtle to measure, this
effect was so pronounced that it was initially impossible to fit
the first cusp crossing at all, until eventually revolution was
included in the fit.

While the two-planet system OGLE-2006-BLG-109Lb,c enabled 
the most dramatic measurement of internal planetary motions,
OGLE-2005-BLG-071 also shows evidence of revolution, despite
its very short, 3-day perturbation.  And, for the reasons
just given, we can expect revolution to be measurable in a significant
fraction of central-caustic events in the future.

\section{Stellar Atmospheres}

Microlensing has proven to be a powerful tool to study stellar
atmospheres in two distinct ways.  First, microlens caustics can
resolve the surfaces of stars better than any other technique,
with the possible exception of transiting planets.  Second,
microlenses can act as a huge magnifying glass to obtain
spectra of stars that would be prohibitively expensive to observe under
ordinary circumstances.

\subsection{Limb Darkening}

\begin{figure}
\plottwo{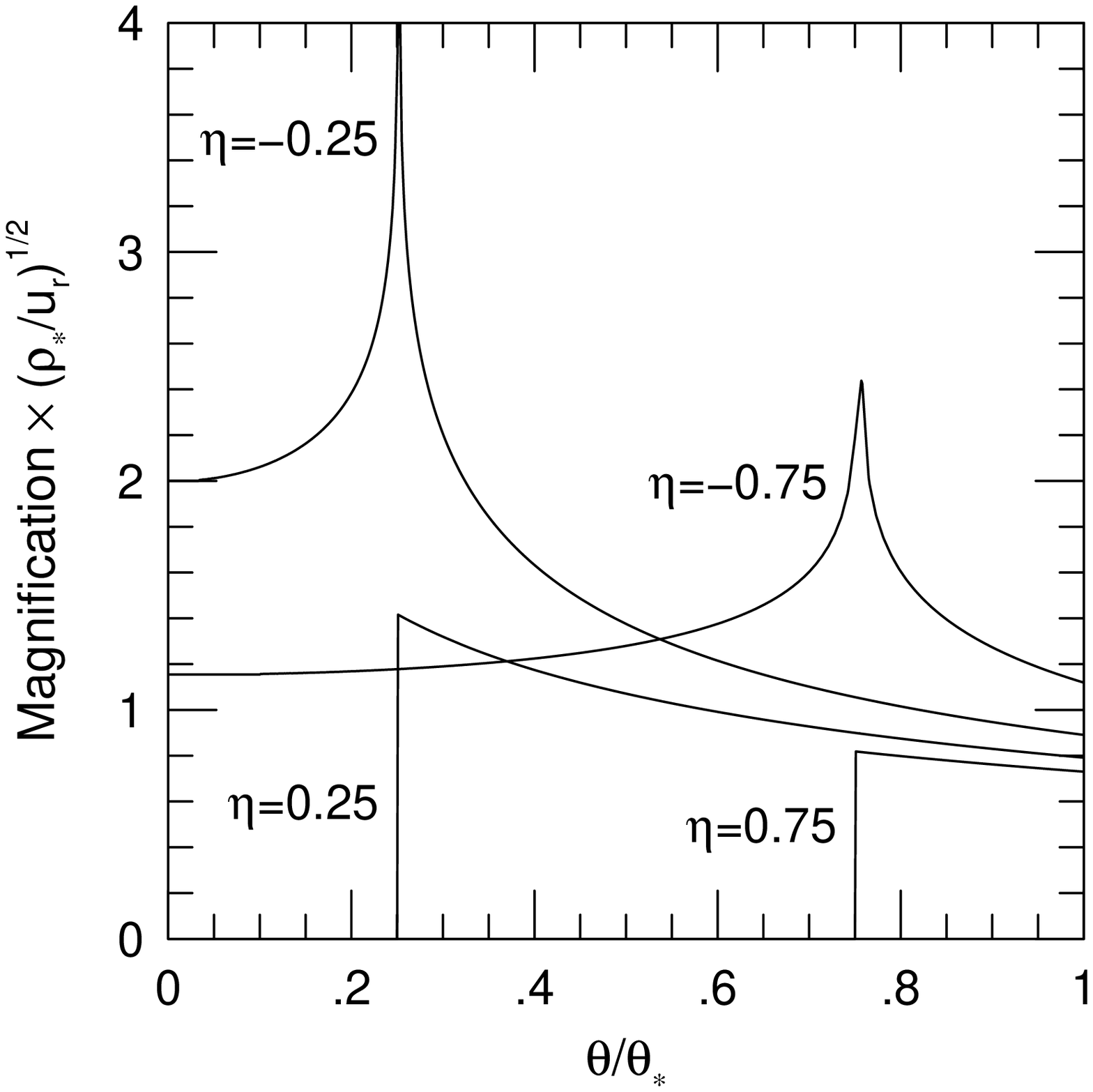}{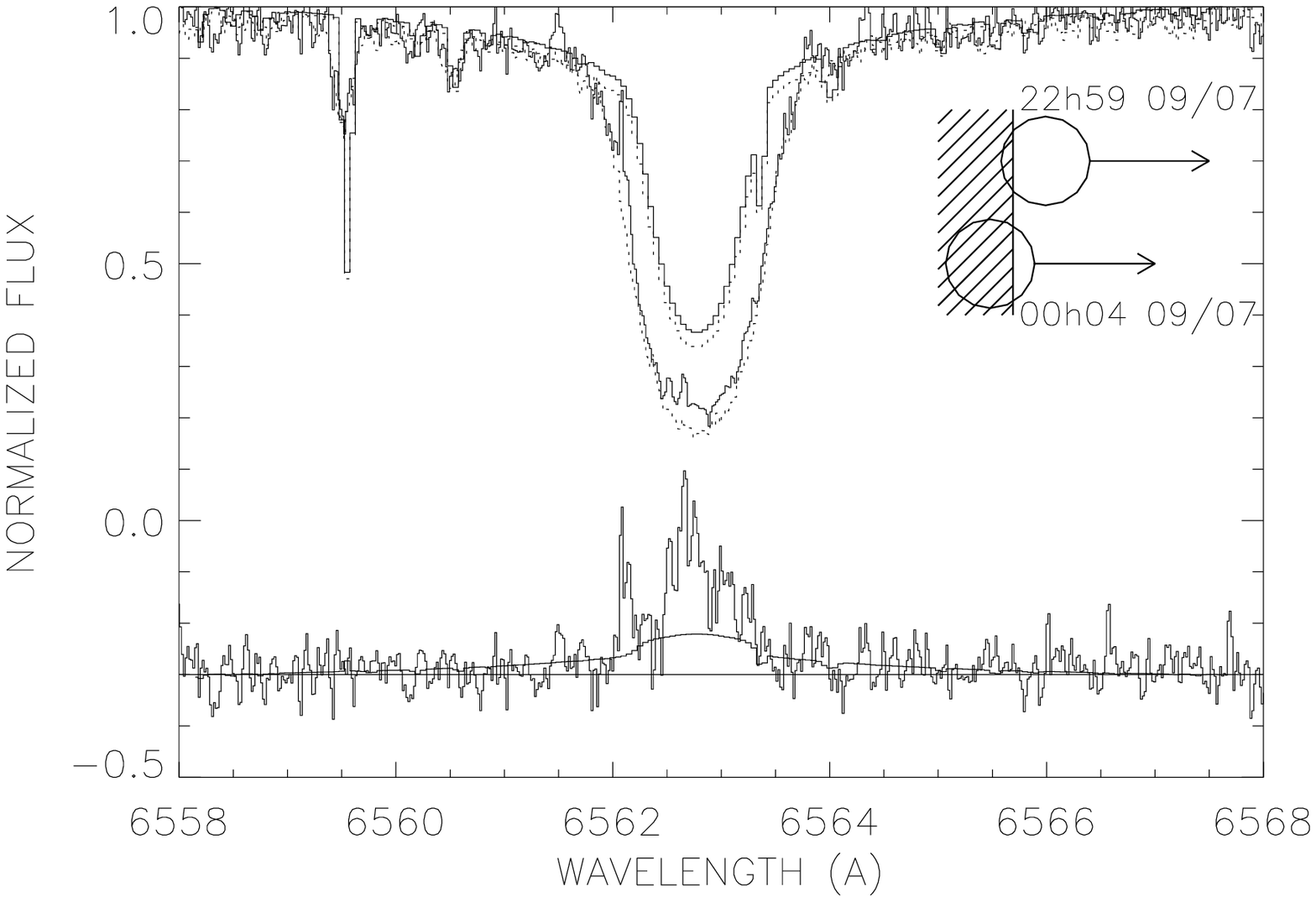}
\caption{\label{fig:ob02069}
Left: Magnification profiles of different stages of a caustic exit.
For example, when the source center
is $\eta=0.75$ source radii outside the caustic,
caustic magnifies the outer 25\% of the source about equally, and
leaves the rest of the source essentially unmagnified.  From \citet{castro01}.
Right: H$\alpha$ profiles of OGLE-2002-BUL-069 when it was just starting
and just finishing its exit (see inset).  The latter shows an
emission bump, probably due to the chromosphere \citep{cassan04}.
}
\end{figure}

Figure \ref{fig:ob02069}a shows the magnification due to a caustic
as a source exits a so-called ``fold caustic'' 
(i.e. a square-root singularity). The surface is assumed to be
radially symmetric (no spots).  Consider first 
the ``$\eta=0.75$'' curve, which describes the magnification profile
when the center of the star is 3/4 of the way past the caustic.
Of course, only the outer 25\% of the star is magnified by the
caustic at all.  Figure \ref{fig:ob02069}a shows that all radii are
magnified about equally.  A photometric series from $\eta=0$ to $\eta=1$
would give a set of box-car convolutions with the radial profile,
permitting straightforward deconvolution of the intrinsic profile.
The $\eta<0$ profiles are more complicated, but do add some additional
information.  (Of course, in addition to the caustic magnification,
there is the underlying non-caustic magnification, which must be
taken into account in the process.  But this is smooth and also
straightforward to model.)\ \ The most spectacular application
of this technique was carried out by \citet{fields03} using PLANET
data on EROS-2005-BLG-5, which had a K-giant source.  
This provided by far the most detailed profile of any star except the
Sun and was also the first (non-solar) confrontation of limb-darkening
models with data.  How did the models do?  They look broadly
similar to the data but do not agree in detail.  In particular,
when K-giant profiles are plotted for a range of temperatures
near that of EROS-2005-BLG-5, they share a ``fixed point'' with
each other and also with the observed profile of the Sun.  But
the deconvolved microlensing profile does not share this ``fixed point''.
Hence, the K-giant models appear to extrapolate from some physics
in the Sun that is not actually shared by K giants.

So far, no atmosphere modelers have risen to this challenge.

\subsection{Chromospheric Spectra}

Two groups obtained spectra of EROS-2000-BLG-5 in an effort to
resolve the source surface simultaneously in spatial and spectral
dimensions \citep{castro01,albrow01}.  In particular, \citet{albrow01}
found about 20\% less H$\alpha$ absorption when, based on the
photometric lightcurve, the source had nearly exited the caustic.
\citet{afonso01} argued that too little of the source was under the
caustic to have such a large effect, and the only plausible explanation 
was that the chromosphere (which was strongly magnified during this
observation) has very strong H$\alpha$ {\it emission}.  Unfortunately,
this conjecture could not be tested in this event because the
\citet{albrow01} spectra were low-resolution, rendering impossible
any identification of separate components to the line.

However, \citet{cassan04} did obtain a high-resolution spectrum of another 
K giant, OGLE-2002-BUL-069,
just as it was ending its exit from a caustic, as well as a comparison spectrum
when it was just beginning its exit.  See Figure~\ref{fig:ob02069}b.
The second spectrum shows a distinct ``bump'' in the H$\alpha$ trough,
confirming strong chromospheric emission.

\subsection{Microlenses as Magnifying Glasses}

\citet{minniti98} published spectroscopic observations of 
microlensed bulge source under the provocative title
``Using Keck I as a 15m Diameter Telescope''.  The idea was that
the source was already magnified by a factor of 2.25, so
these observations were equivalent to using a 15m telescope
at the same exposure time.  Now, events remain magnified by
a factor 2 for a week or two, so with some modest planning one
could arrange to cut down on long exposures considerably this way.
But when planetary microlensers began concentrating on high-mag
events,  much more dramatic improvements became possible.

\begin{figure}
\plottwo{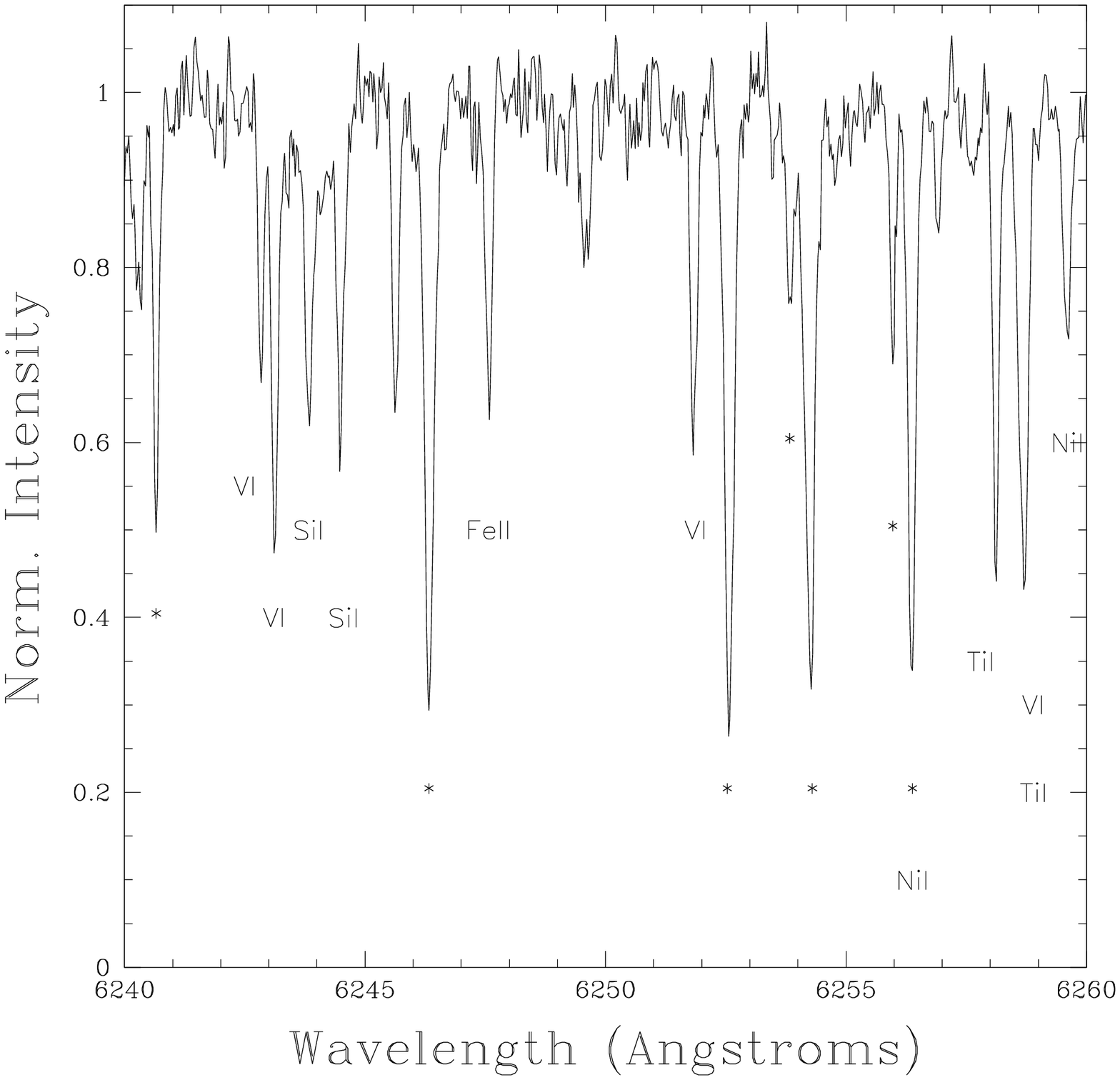}{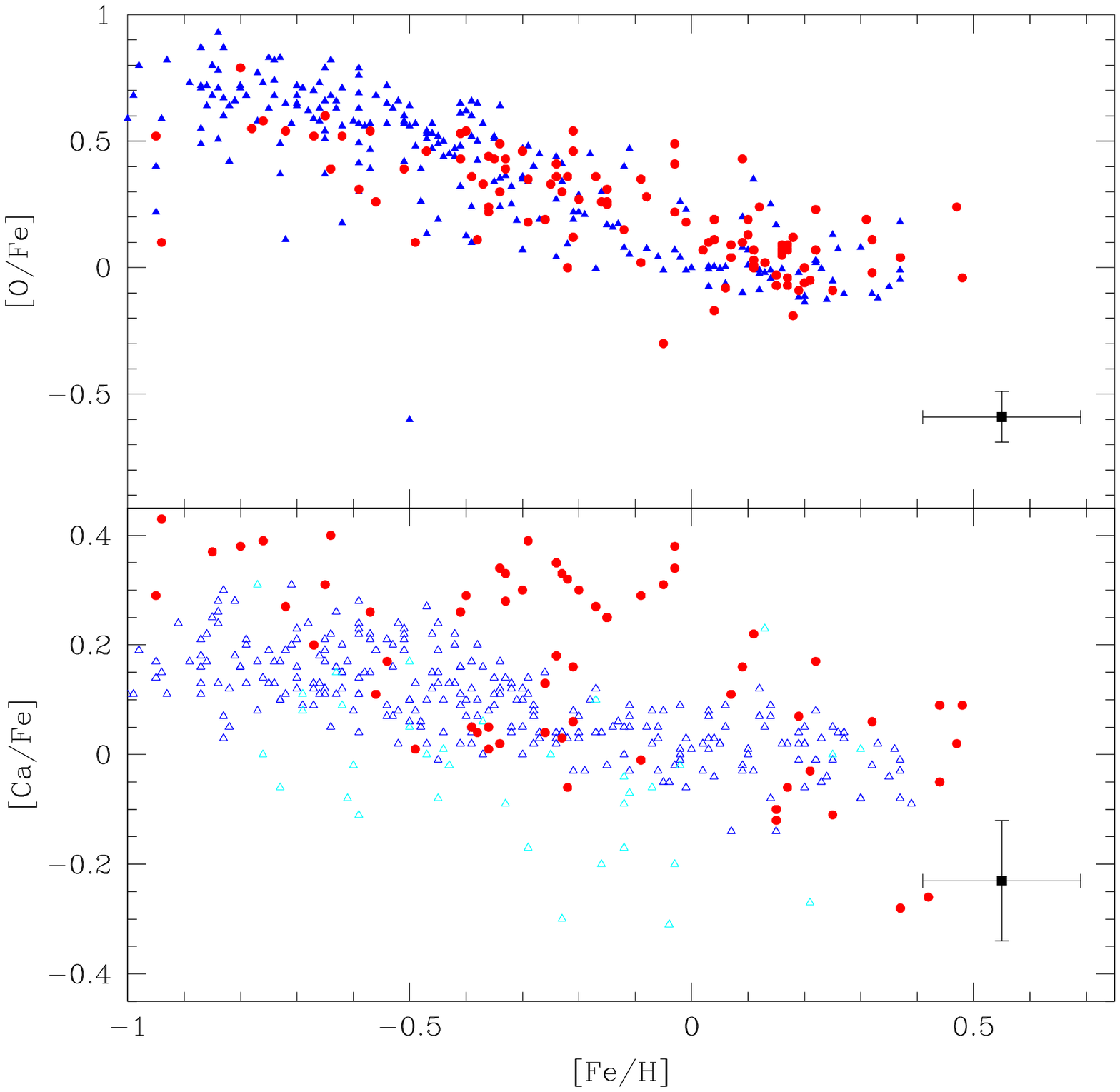}
\caption{\label{fig:ob06265}
Using Keck as a 115m telescope.
Left: 15 minute Keck spectrum of OGLE-2006-BLG-265 when it was
magnified by $A=130$.
Right: Abundance measurements derived by \citet{johnson07}.
The first bulge dwarf with a very high quality spectrum proves
to be one of the most metal-rich bulge stars.  
}
\end{figure}

The first example of this occurred in a quite unplanned way.
Avishay Gal-Yam was at Keck when (as a member of $\mu$FUN)
he received a flurry of emails urgently requesting {\it photometric}
observations of OGLE-2006-BLG-265, which eventually reached
magnification $A=230$.  He decided to get a 15 minute spectrum
(at $A=130$, it turned out), thus using Keck as a 115m telescope
on this $I=19.4$ star!
This was by far the best spectrum
of a bulge dwarf to that time.  See Figure \ref{fig:ob06265}a.  
Once again, the initiative of the observer proved crucial.
\citet{johnson07} analyzed this spectrum and found a G-dwarf
with [Fe/H]$=0.55$, one of the most metal rich stars yet observed.
See Figure \ref{fig:ob06265}b.  

\begin{figure}
\plottwo{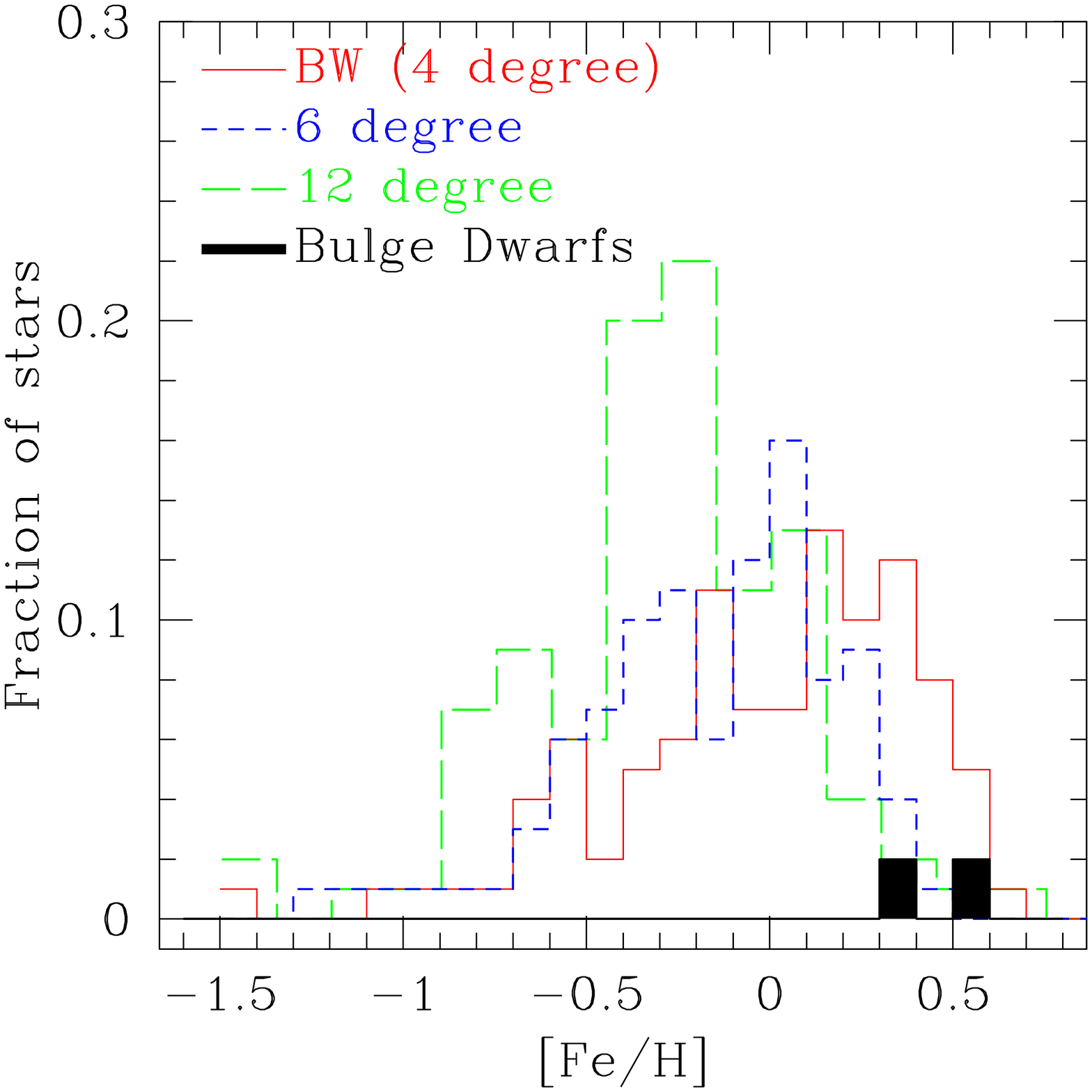}{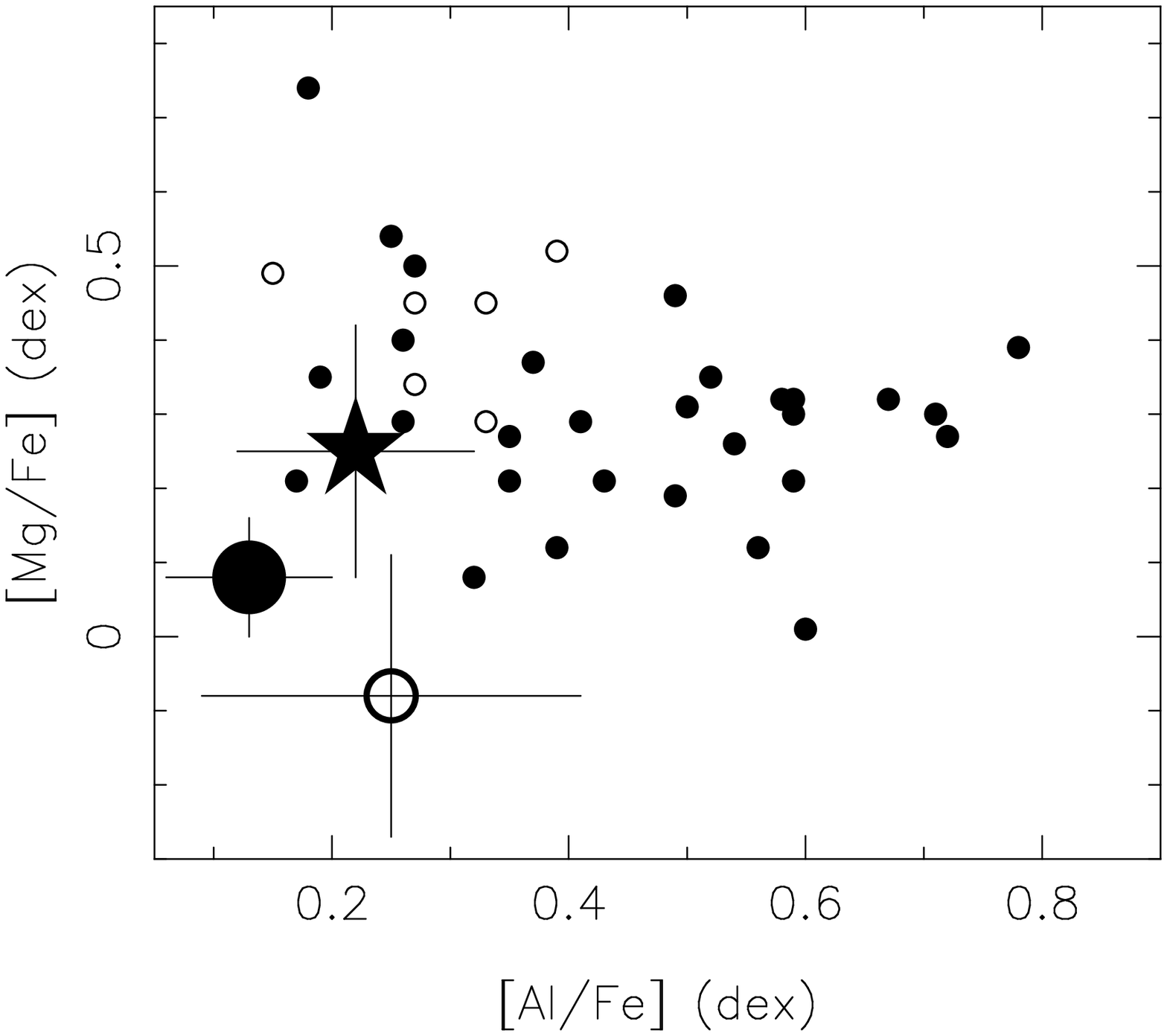}
\caption{\label{fig:ob07349}
Left: Iron abundances of first two highly-magnified bulge dwarfs ({\it black}),
OGLE-2006-BLG-265 (see Fig.~\ref{fig:ob06265}) and MOA-2006-BLG-099
\citep{johnson08} compared to those of bulge giants (histograms).
A third dwarf, OGLE-2007-BLG-349 \citep{cohen08}, also has [Fe/H]=+0.5,
making the dwarf and giant distributions inconsistent at $4\times 10^{-5}$.
Right: [Mg/Fe] and [Al/Fe] ratios of the three highly magnified dwarfs 
({\it large symbols}) compared to those of bugle giants \citep{cohen08}.
}
\end{figure}

These results inspired a somewhat more systematic effort to
obtain such spectra as part of the ``normal'' frenetic activity
that surrounds high-mag events.  Scott Gaudi obtained another
spectrum a few weeks later of MOA-2006-BLG-099 \citep{johnson08}, and another
substantially higher S/N spectrum of OGLE-2007-BLG-349 was obtained by 
Judy Cohen the next year \citep{cohen08}.  These two dwarfs are also
iron rich, and a KS test (probability $4\times 10^{-5}$)
shows that these metallicities are not drawn from the same distribution
found for giants.  See Figure~\ref{fig:ob07349}.
The most likely explanation is that metal-rich dwarfs
blow off their envelopes before they can become evolved giants, so
giant stars are not representative of the underlying population.

\section{Coming Full Circle: Domestic Microlensing Event}

\citet{einstein36} famously dismissed microlensing in the very paper
he introduced it:  ``Some time ago R.W.\ Mandl paid me a visit
and asked me to publish the results of a little calculation, which
I had made at his request $\ldots$ there is no great chance of
observing this phenomenon.''  In fact, Rudi Mandl, a Czech electrical engineer, 
after perhaps failing to gain Einstein's attention by mail, obtained
``a small sum of money'' from the Science Service
to come to Princeton to pester Einstein in person.
Einstein already knew, or thought he knew, that microlensing was 
unobservable because he had already worked out the magnification
and cross section in 1912 \citep{renn97}.  Hence, as his private
remarks to the editor of {\it Science} reveal, he was actually
far more dismissive of this idea than even his article indicated:
``Let me also thank you for your cooperation with the little publication,
which Mister Mandl squeezed out of me.  It is of little value, but
it makes the poor guy happy.''

Why was Einstein so down on microlensing?  One reason appears sound.
In 1936, photographic catalogs went to about $V=12$, about the limit
of the Tycho-II catalog.  So there would have been of order
2 million stars, the giants among which would typically be at about
2 kpc.  It is straightforward to work out that the optical depth for these
stars is $\tau \sim 10^{-8}$ and that they have an event rate 
$\Gamma\sim 10^{-7}\,\rm yr^{-1}$.  Hence, even if all these stars
were monitored continually with a precision much better than 0.3
mag (a complete impossibility in Einstein's day), there would be
only 1 event every 10 years.

In fact,  it is unlikely that Einstein ever did this calculation.
For one thing, he evaluates what we would call the Einstein radius
as ``a few light seconds'', whereas it is more like a few hundred
light seconds, meaning that he was discouraged from doing a detailed
calculation before he got to this stage.  But for another, one
gains the definite impression from his article that he was 
thinking of microlensing as a static, not dynamic phenomenon.  
He says (in our notation) that $u$ ``must be small compared to [unity],
[to produce] an appreciable increase of the apparent brightness'' 
of the source.  This implies that he considered the phenomenon to
be unobservable unless the magnification were very high, much greater
than unity.  Since low amplitude variables were known in Einstein's
day, this must mean that he was not thinking about
{\it microlensing events}, but rather thought that for
a microlensed star to be noticed, it would have to be anomalously
bright (e.g., for its color).  
And recognizing microlensing events by this path would 
indeed be extraordinarily difficult, even today.
It appears that it was \citet{russell37} who thought of the idea
of microlensing {\it events}, albeit in a context rather different
from the ones we observe today.  So, while the term ``Einstein ring''
does reflect the real history of this subject, perhaps it
would have been more appropriate to refer to the ``Russell timescale''.

\begin{figure}
\plottwo{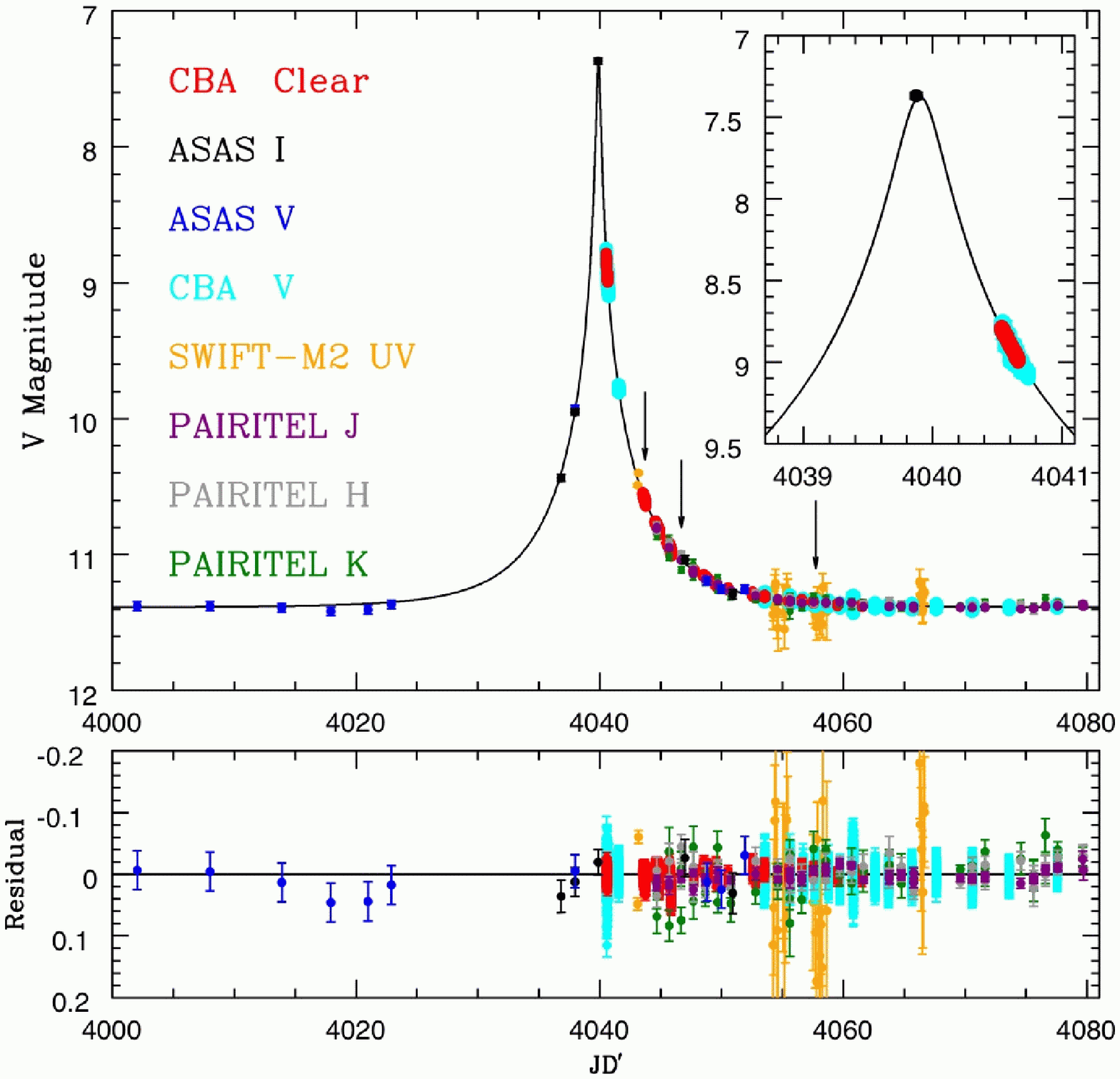}{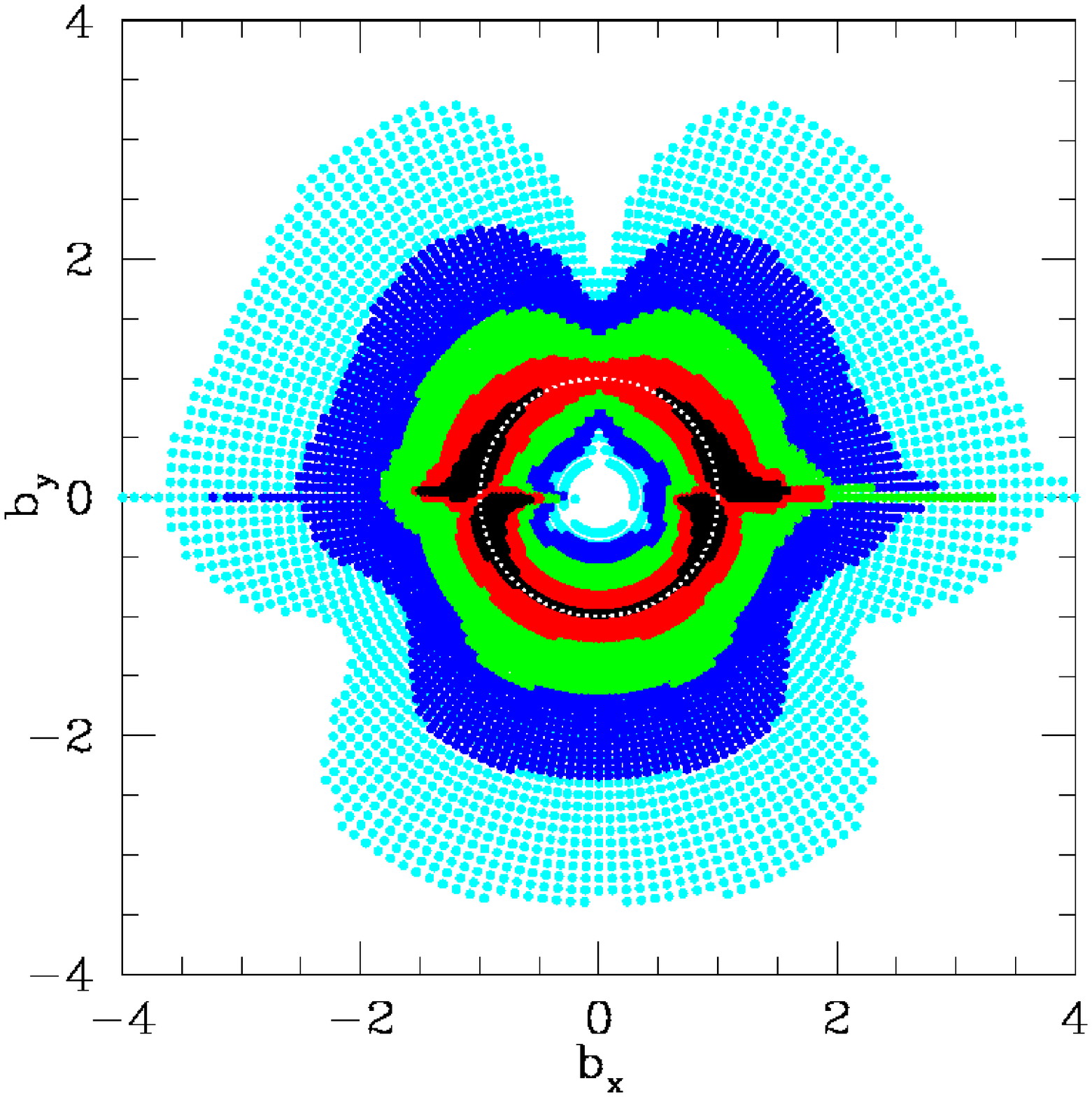}
\caption{\label{fig:domestic}
Domestic Microlensing Event.
Left: Light curve of microlensing event of a nearby $V=11$ A star,
discovered by amateur Akihiko Tago  and monitored on its fall by Joe Patterson's
Center for Backyard Astrophysics.  Grzegorz Pojma\'nski's ASAS all-sky
monitoring, recovered after the event, proved crucial in 
demonstrating that the lightcurve is symmetric, and so is almost certainly
microlensing.
Right: Planet sensitivity of this event assuming that ASAS observations
had been analyzed in time to issue a microlensing alert, thus allowing
the event to be monitored intensively over peak.  Contours vary from
$q=10^{-5}$ ({\it black}) to $q=10^{-3}$ ({\it cyan}).  Planet-star
separation is in units of the Einstein radius.
From \citet{gaudi08b}
}
\end{figure}

A recent observation of a ``domestic microlensing event'' calls into question
Einstein's dismissal, even judged on its own terms. And here again,
Bohdan played a role, albeit indirect.
Akihiko Tago, a Japanese
amateur who has been scanning the sky for 40 years for novae and comets,
noticed that a $V=11$ A star had suddenly brightened by 4 magnitudes.
He issued an alert which was picked up by Joe Patterson's 
Center for Backyard Astrophysics (CBA), a network of amateurs and
professionals dedicated to variable phenomena.  After Joe had ruled
out all other explanations, he concluded that  the lightcurve could
only be microlensing and sent the data to Scott Gaudi and me.
I promptly told him it
could not be microlensing for two reasons.  First, by the argument that
Einstein either made, or might have made, such events are too rare.
Indeed the above calculation was for the rate of events with 
impact parameters $u_0<1$, i.e., all source trajectories that cross
any part of the Einstein ring.  However, this event, if it indeed were
microlensing, would have been magnified 40 times and so would have
been 40 times rarer.  But second, even combining the discovery data and
the CBA data, only the falling part of the lightcurve was available.
It is well known that the falling lightcurves of novae and other 
eruptive variables can look like microlensing, but are easily
discriminated by the asymmetry between their rise and fall.
Even if this A star was not some known type of variable, without
a rising lightcurve, it was more likely to be an unknown variable
than microlensing, although lack of Xray emission and 
emission features in the optical spectrum did tend to weigh against
an eruptive variable.

Enter Grzegorz Pojma\'nski and his ASAS project, which as he relates
in this volume, was nurtured and encouraged by Bohdan.  ASAS was
already functioning in the south, but had only begun test observations
in the north. When contacted, Pojma\'nski found that ASAS observations
covered both the rise and fall of the event, and indeed one
observation right at peak.  See Figure \ref{fig:domestic}a.

These proved two things.  First the event is symmetric and so almost
certainly is microlensing \citep{fukui07,gaudi08b}.  
Second, it could not have been recognized
as microlensing just based on observations (by amateurs or professionals)
that could have been made in Einstein's day.  Automated observations
of large parts of the sky are required.

While microlensing events of stars at 1 kpc probably really are quite
rare, relatively nearby events at 4 kpc occur several times per year.
At the end of his life, Bohdan was thinking about next generation
wide-field surveys that could detect these.  If detected and
publicized before peak, such events could open a new avenue of planet
detection.  Figure \ref{fig:domestic}b shows the sensitivity
to planets (including Earth-mass planets {\it black}) of hypothetical 
observations of the event at left, assuming that it had been alerted in time
to densely monitor the peak.

\section{Conclusions}

The microlensing surveys that began in the 1990s are directly traceable
to Bohdan's influence, inspiring the Magellanic Cloud surveys with
his 1986 article and directly helping to initiate and guide OGLE.
As advocated by \citet{pac86}, these searches began by looking for
dark matter, but Bohdan began immediately to push microlensing in
new directions, particularly planets and Galactic structure.
Two decades later, microlensing has become an incredibly powerful
tool and an incredibly rich subject.


\acknowledgements 
I thank Rich Gott for valuable insights into the early history of microlensing,
and Scott Gaudi for a careful review of the manuscript.
This work was supported in part by grant AST-042758 from the NSF.


\end{document}